\documentclass[journal]{IEEEtran}

\ifCLASSINFOpdf

\else

\fi

\usepackage{calc}
\usepackage{graphics,color,wrapfig,hyperref}
\usepackage{color}
\usepackage{amsthm}
\usepackage{amsmath,amssymb}
\usepackage{amsfonts}
\usepackage{array}
\newcolumntype{P}[1]{>{\centering\arraybackslash}p{#1}}
\usepackage{multirow}
\usepackage{graphicx}
\usepackage{standalone}
\usepackage{booktabs}
\usepackage{algpseudocode}
\usepackage{algorithm2e}
\let\oldnl\nl
\newcommand{\nonl}{\renewcommand{\nl}{\let\nl\oldnl}}
\usepackage{makecell}
\usepackage[bottom]{footmisc}
\usepackage{enumitem}
\theoremstyle{definition}
\usepackage{tikz}
\usepackage{xcolor}

\makeatletter
\renewcommand{\@algocf@capt@plain}{above}
\makeatother
\RestyleAlgo{ruled}
\LinesNumbered

\usepackage{enumitem}
\theoremstyle{definition}

\usepackage[lofdepth,lotdepth]{subfig}

\newcommand{\subparagraph}{}
\usepackage{titlesec}
\titlespacing\section{0pt}{12pt plus 4pt minus 2pt}{0pt plus 2pt minus 2pt}
\titlespacing\subsection{0pt}{12pt plus 4pt minus 2pt}{0pt plus 2pt minus 2pt}

\newcommand{\calA}{\mathcal{A}}

\newcommand{\calP}{\mathcal{P}}
\newcommand{\calX}{\mathcal{X}}
\newcommand{\calG}{\mathcal{G}}

\newcommand{\bfg}{\mathbf{g}}

\newtheorem{observation}{Observation}
\newtheorem{lemma}{Lemma}

\newtheorem{definition}{Definition}

\newtheorem{example}{Example}

\newcounter{relctr} 
\everydisplay\expandafter{\the\everydisplay\setcounter{relctr}{0}} 

\newcommand\labelrel[2]{%
  \begingroup
    \refstepcounter{relctr}%
    \stackrel{\textnormal{(\alph{relctr})}}{\mathstrut{#1}}%
    \originallabel{#2}%
  \endgroup
}
\AtBeginDocument{\let\originallabel\label} 

\begin{document}

\title{Resolvable Designs for Speeding up Distributed Computing}

\author{\IEEEauthorblockN{Konstantinos Konstantinidis and Aditya Ramamoorthy}\\
%
\thanks{ This work was supported in part by the
National Science Foundation (NSF) under grants CCF-1718470 and CCF-1910840. The material in this work has appeared in part at the 2019 IEEE International Symposium on Information Theory and the 2018 IEEE Global Telecommunications Conference (GLOBECOM). The authors are with the Department of Electrical and Computer Engineering, Iowa State University, Ames, IA, 50011 USA. e-mail: \{kostas@iastate.edu, adityar@iastate.edu\}
}
}

\maketitle

\begin{abstract}

Distributed computing frameworks such as MapReduce are often used to process large computational jobs. They operate by partitioning each job into smaller tasks executed on different servers. The servers also need to exchange intermediate values to complete the computation. Experimental evidence suggests that this so-called Shuffle phase can be a significant part of the overall execution time for several classes of jobs. Prior work has demonstrated a natural tradeoff between computation and communication whereby running redundant copies of jobs can reduce the Shuffle  traffic load, thereby leading to reduced overall execution times. For a single job, the main drawback of this approach is that it requires the original job to be split into a number of files that grows exponentially in the system parameters. When extended to multiple jobs (with specific function types), these techniques suffer from a limitation of a similar flavor, i.e., they require an exponentially large number of jobs to be executed. In practical scenarios, these requirements can significantly reduce the promised gains of the method. In this work, we show that a class of combinatorial structures called resolvable designs can be used to develop efficient coded distributed computing schemes for both the single and multiple job scenarios considered in prior work.  We present both theoretical analysis and exhaustive experimental results (on Amazon EC2 clusters) that demonstrate the performance advantages of our method. For the single and multiple job cases, we obtain speed-ups of 4.69x (and 2.6x over prior work) and 4.31x over the baseline approach, respectively.

\end{abstract}

\begin{IEEEkeywords}
MapReduce, data-intensive computing, coded multicasting, communication load, TeraSort, aggregate functions, distributed learning.
\end{IEEEkeywords}

%
\IEEEpeerreviewmaketitle

\section{Introduction}
\label{sec:intro}

In recent years, there has been a surge in the usage of various cluster computing frameworks such as MapReduce, Hadoop and Spark. The era of big data analytics whereby a large amount of data needs to be processed in a fast manner has fueled this growth. In these systems, the data set is usually split into disjoint files stored across the servers. The computation takes place in three steps. In the \emph{Map} step, the servers process the input files to generate \emph{intermediate values} having the form of (key, value) pairs. In the next \emph{Shuffle} step, the intermediate pairs are exchanged between the servers. In the final \emph{Reduce} step, each server computes a set of output functions defined based on the keys. Henceforth, we refer to this as the MapReduce framework.

The MapReduce framework has proven to be quite versatile and large scale clusters in industry and academia routinely process terabytes of data using this approach. It is a protocol well suited for several applications; it fits the computation of functions which are useful for machine learning \cite{compressed_CDC}, e.g., in deep residual learning for image recognition \cite{he_resnet}. Prakash et al. \cite{saurav_g_analytics} have adapted the general MapReduce framework to graph analytics where computation at each vertex of the graph requires data only from the neighboring vertices.
It is important to note that the framework intertwines computation and communication. Specifically, multiple workers allow for parallel computation; yet data needs to be exchanged between them to complete the processing of the job. The terms \emph{servers} and \emph{workers} will be used interchangeably throughout the text.

A typical MapReduce implementation splits the overall job into a number of equal-sized (or approximately equal-sized) tasks and assigns a single task to each server. However, for many classes of jobs, extensive experimental results have shown that in such implementations the Shuffle phase can be quite expensive and dominates the overall execution time \cite{Chowdhury_etal11}. There have been several papers \cite{Chowdhury_etal11, GuoRZ13, Verma2013, EZZELDIN_KARMOOSE_FRAGOULI_2017}, on the impact of the Shuffle phase on the overall execution of a MapReduce job and corresponding work on alleviating it. These effects have been reported in the work of Guo et al. \cite{GuoRZ13} on Shuffle-heavy operations such as SelfJoin, TeraSort and RankedInvertedIndex. Distributed graph analytics also suffer from long communication phases as observed in \cite{saurav_g_analytics} and \cite{Chen_graph14}.


The CDC scheme in \cite{LiMA16} (see also \cite{compressed_CDC}) showed an interesting information theoretic perspective on trading off computation vs. communication. The basic technique they suggest is to introduce redundancy in the computation, i.e., execute multiple copies of a given Map task at different servers and use coded transmissions to reduce the amount of data exchanged during the Shuffle phase. The servers use locally available intermediate values in order to decode the received messages and compute their output functions. Their work for a general MapReduce system characterizes and matches the information-theoretic lower bound on the minimum communication load under certain assumptions.

In this work, we demonstrate that in practical scenarios, the original scheme in \cite{LiMA16} and \cite{compressed_CDC} require significantly higher shuffling time than the theoretical prediction. This stems from the requirement, e.g., that a given job needs to be split into a large number of small tasks in \cite{LiMA16} and we show that it has detrimental effects on the performance of the method. In this work, we present a technique based on using combinatorial structures known as resolvable designs for exploring the computation vs. communication tradeoff within distributed computation and demonstrate its advantages.

\section{Background, Related Work and Summary of Contributions}

Ahmad at al. \cite{Ahmad_2014} introduced ``ShuffleWatcher", a MapReduce scheduler that reduces throughput and job completion time. The scheme replicates Map tasks and delays or elongates a job's communication time depending on the network load. Other related work on this topic has been published in \cite{Cao_2016} which considers a model of MapReduce executed on a multi-core machine and proposes a topology-aware architecture to expedite data shuffling. Wang et al. \cite{Wang_2013} present an algorithm that finds the optimal placement and jointly optimizes Map and Shuffle time.

To our best knowledge \cite{LiMA16}, \cite{LiMAAllerton15} and \cite{songze_terasort} were the first to rigorously examine the MapReduce framework within the computation vs. communication tradeoff. Their work defines appropriate notions of computation and communication loads within MapReduce. Their key finding is that the judicious usage of coded transmissions in the Shuffle phase can significantly reduce the communication load. Compared to a baseline scheme, their algorithm splits the original job more finely into a certain number of Map tasks and redundantly assigns each of them to multiple workers. Nonetheless, their work requires splitting the job into a very large number of files or tasks. This limitation hurts their scheme in a number of different ways; the most immediate one is that they require extremely large data sets as the cluster size scales. Their proposed method also has to form many shuffling groups of servers communicating in the Shuffle phase. For each group, each participating server will initially form an encoded packet to transmit to the rest of the group; all these packets are stored in the memory of the server. As a result, their approach suffers from a significant overhead in encoding time accounting for all groups.

The idea of Compressed Coded Distributed Computing (CCDC), presented in \cite{compressed_CDC}, applies to scenarios where the underlying functions being computed can be aggregated. Examples of such functions include, e.g., Average(), Count(), Max(), Min(), Median(), Mode(), Range() and Sum(). This kind of computation is predominant in machine learning (e.g., ImageNet classification \cite{he_resnet} and stochastic gradient descent \cite{tandon_gradient}). Another scenario is matrix-vector multiplications that are performed during the forward and backward propagation in neural networks (\emph{cf.} \cite{Dally_nips_tutorial}). This so-called \emph{compression technique} was initially investigated in \cite{DeanG08} by means of a ``combiner function" which merges intermediate values with the same key computed from different Map functions. This allows for a potential reduction in network traffic as intermediate values can be aggregated before transmission in the Shuffle phase. Interestingly, \cite{compressed_CDC} requires the number of jobs being processed simultaneously to be very large. This can also be a restrictive assumption in practice.

The recent work of Woolsey et al. in \cite{WOOLSEY_CHEN_MINGYUE_2018} introduces a scheme to handle the case when each Reduce function is computed by $s>1$ workers by utilizing a hybercube structure which controls the allocation of Map and Reduce tasks. Their work is motivated by distributed applications that require multi-round Map and Reduce computations. Another approach that re-examines the computation-communication tradeoff from an alternate viewpoint has been investigated in \cite{EZZELDIN_KARMOOSE_FRAGOULI_2017}. In this case, the assumption is that a server does not need to process all locally available files and storage constraints do not necessarily imply computation constraints. A lower bound on the \emph{computation load} and a heuristic scheme were derived. In \cite{Song_Srinivasavaradhan_Fragouli_2017}, the authors propose a scheme which gives each server access to a random subset of the input files and not all Reduce functions depend on the entire data set. 


\subsection{Summary of contributions}

As discussed above both \cite{compressed_CDC} and \cite{LiMA16} require a certain problem dimension to be very large. In particular, \cite{LiMA16} considers a single job and requires it to be split into a number of tasks that grows exponentially in the problem parameters. On the other hand \cite{compressed_CDC} considers functions that can be aggregated but requires the number of jobs being processed simultaneously to grow exponentially. Our work builds on our initial work in \cite{konstantinidis_ramamoorthy_globecom} and \cite{kostasR19} and makes the following contributions.

\begin{itemize}
\item We demonstrate a natural link between the problem of reducing MapReduce Shuffle traffic and combinatorial structures known as resolvable designs \cite{DRSCDCA}, which in turn can be easily generated from linear error correcting codes.
\item For the single-job case, our resolvable design based scheme significantly reduces the number of files compared to \cite{LiMAAllerton15}, \cite{LiMA16} and \cite{songze_terasort}. As compared to an uncoded scheme, CDC in \cite{LiMA16} reduces the shuffle phase load by a factor of $r$ if each task is executed on $r$ workers. In contrast, our technique reduces the Shuffle phase load by a factor of $r-1$, but requires much fewer files. It turns out that in practice our method has a higher gain. For instance, our experiments (\emph{cf.} Section \ref{sec:results_one_job}) have an overall speedup of $3.01\times$ compared to \cite{songze_terasort} where the procedure of \cite{LiMA16} has been applied to a sorting algorithm.
\item 
    
    For the multi-job case we seek a method that combines the benefits of the coding-theoretic ideas employed in the single-job case and the fact that the functions being computed are amenable to aggregation. A simple strategy in the multi-job case would be to simply use the single-job ideas in a sequential manner. However, our work shows that a careful assignment of jobs and tasks to the worker nodes and exploiting the aggregation property can reduce the Shuffle load significantly. In particular, our work requires much fewer jobs than CCDC in \cite{compressed_CDC}, while enjoying the exact same Shuffle phase load.

\item For both problems we present  exhaustive experimental comparisons on Amazon EC2 clusters with prior work that demonstrate the efficacy of our method. The code for our techniques is publicly available at \cite{KKCT} and \cite{compressed_repo}.
\end{itemize}

Existing distributed frameworks (\emph{cf.} Hadoop/Spark) typically use redundancy for a different purpose (e.g., fault tolerance) while we use it to reduce the Shuffle traffic. Our work has not been proposed as a variant or incremental fix to these frameworks. It is demonstrating that a coding-theoretic viewpoint has the potential to yield great dividends within the Shuffle phase traffic reduction. Our implementations use C++ and MPI for a head-to-head comparison with the work of \cite{LiMA16}. Our approaches are applicable to problems requiring long communication phases where a decrease in Shuffle time can significantly offset an increase in Map time due to redundancy. The translation and/or adaptation of our approaches into protocols that are used in practice is not the focus of our work but we hope that this theoretical/numerical evidence spurs more research in this area.

\section{Preliminaries}
\label{sec:scheme_one_job}

\subsection{Primer on resolvable designs}
\label{sec:primer_resolvable}
We begin with some basic definitions from combinatorial design theory \cite{DRSCDCA} that we need for specifying our protocols.
\theoremstyle{definition}
\begin{definition}
A \emph{design} is a pair ($X$, $\mathcal{A}$) consisting of
\begin{enumerate}
\item a set of elements (\emph{points}), $X$, and
\item a family $\mathcal{A}$ (i.e. multiset) of nonempty subsets of $X$ called \emph{blocks}, where each block has the same cardinality.
\end{enumerate}
\end{definition}

\begin{definition}
A subset $\mathcal{P}\subset \mathcal{A}$ in a design $(X,\mathcal{A})$ is said to be a \emph{parallel class} if for $X_i\in\mathcal{P}$ and  $X_j\in\mathcal{P}$ with $i\neq j$ we have $X_i\cap X_j=\emptyset$  and $\cup_{\{j:X_j\in \mathcal{P}\}}X_j=X$. A partition of $\mathcal{A}$ into several parallel classes is called a \emph{resolution} and $(X,\mathcal{A})$ is a \emph{resolvable design} if $\mathcal{A}$ has at least one resolution.
\end{definition}

\begin{example}
\label{eg:subsets of_four}
Let $X=\{1,2,3,4\}$ and  $\mathcal{A}=\{\{1,2\},\{3,4\},\{1,3\},\{2,4\},\{1,4\},\{2,3\}\}$. The $(X, \calA)$ forms a resolvable design with the following parallel classes
\begin{gather*}
\mathcal{P}_1=\{\{1,2\},\{3,4\}\}, \mathcal{P}_2=\{\{1,3\},\{2,4\}\}\text{ and }\\
\mathcal{P}_3=\{\{1,4\},\{2,3\}\}.
\end{gather*}

\end{example}
It turns out that there is a systematic procedure for constructing resolvable designs, where the starting point is an error correcting code. We explain this procedure below.

Let $\mathbb{Z}_q$ denote the additive group of integers modulo $q$ \cite{lincostello}. The generator matrix of an $(k,k-1)$ single parity-check (SPC) code over $\mathbb{Z}_q$\footnote{We emphasize that this construction works even if $q$ is not a prime, i.e., $\mathbb{Z}_q$ is not a field.} is defined by
\begin{equation}
\mathbf{G}_{SPC}=
\left[
\begin{array}{ccc|c}
& & &1\\
&\huge \mathbf{I}_{k-1}& &\vdots\\
& & &1
\end{array}
\right].
\label{eqn:spcgen}
\end{equation}
This code has $q^{k-1}$ codewords which are given by $\mathbf{c} = \mathbf{u} \cdot \mathbf{G}_{SPC}$ for each possible message vector $\mathbf{u}$. The code is systematic so that the first $k-1$ symbols of each codeword are the same as the symbols of the message vector. The $q^{k-1}$ codewords $\mathbf{c}_i$ computed in this manner are stacked into the columns of a matrix $\mathbf{T}$ of size $k\times q^{k-1}$, i.e.,
\begin{equation}
\mathbf{T}=[{\mathbf{c}}_1^T,{\mathbf{c}}_2^T,\cdots,{\mathbf{c}}_{q^{k-1}}^T].
\label{eqn:T}
\end{equation}
The corresponding resolvable design is constructed as follows. Let $X_{SPC} = [q^{k-1}]$ (we use $[n]$ to denote the set $\{1,2,\dots,n\}$ throughout) represent the point set of the design.
We define the blocks as follows. For $0 \leq l \leq q-1$, let $B_{i,l}$ be a block defined as $B_{i,l}=\{j: \mathbf{T}_{i,j}=l\}$.

The set of blocks $\mathcal{A}_{SPC}$ is given by the collection of all $B_{i,l}$ for $1 \leq i \leq k$ and $0 \leq l \leq q-1$ so that $|\mathcal{A}_{SPC}| = kq$. The following lemma (proved in \cite{TangR18}) shows that
this construction always yields a resolvable design. 

\theoremstyle{lemma}
\begin{lemma}
The above scheme always yields a resolvable design $(X_{SPC},\mathcal{A}_{SPC})$ with $X_{SPC}=[q^{k-1}]$, $|B_{i,l}|=q^{k-2}$ for all $1\leq i\leq k$ and $0\leq l\leq q-1$. The parallel classes are analytically described by $\mathcal{P}_i=\{B_{i,l}:0\leq l\leq q-1\}$, for $1\leq i\leq k$.
\end{lemma}

\begin{table*}[!t]
\centering
\caption{Proposed placement scheme for Example \ref{eg:N4_example}.}
\resizebox{1.25\columnwidth}{!}{
\begin{tabular}{ |c|c|c|c|c|c|c| }
\hline
Parallel class&\multicolumn{2}{c|}{$\mathcal{P}_1$}&\multicolumn{2}{c|}{$\mathcal{P}_2$}&\multicolumn{2}{c|}{$\mathcal{P}_3$}\\
\hline
Server&$U_1$&$U_2$&$U_3$&$U_4$&$U_5$&$U_6$\\
\hline
Mapped files&$w_1,w_2$&$w_3,w_4$&$w_1,w_3$&$w_2,w_4$&$w_1,w_4$&$w_2,w_3$\\
\hline
\end{tabular}
}
\label{table:placement_eg_N4_example}
\end{table*}

\begin{example}
\label{ex:classicspc}
The generator matrix of this $(3,2)$ SPC code over $\mathbb{Z}_2$ (binary), i.e., for $k=3$ and $q=2$ is given by
$\mathbf{G}_{SPC}=
\begin{bmatrix}\mathbf{I}_{2\times2}&\mathbf{1}_{2\times1}\end{bmatrix}.
$
The matrix $\mathbf{T}$ can be obtained as
\begin{equation*}
\mathbf{T}=[{\mathbf{c}}_1^T,{\mathbf{c}}_2^T,{\mathbf{c}}_3^T,{\mathbf{c}}_4^T]
=\begin{bmatrix}
0&0&1&1\\
0&1&0&1\\
0&1&1&0
\end{bmatrix}.
\end{equation*}



It can be observed, e.g., that $B_{1,0} = \{1,2\}$ and $B_{1,1} = \{3,4\}$ so that they form a parallel class. In fact, this construction returns the resolvable design considered in Example \ref{eg:subsets of_four}.
\end{example}

\subsection{Main Shuffling Algorithm}
Throughout the paper we specify the Shuffle phases by means of various coded transmissions. The following lemma is repeatedly used in the sequel; the proof is in the Appendix.


\begin{lemma}
\label{lemma:shufffling_lemma}
Consider a group of $k$ servers $G=\{U_1,\dots,U_k\}$ with the property that every server in $G\setminus\{U_{\ell}\}$, stores a chunk of data of size $B$ bits, denoted $\mathcal{D}_{[\ell]}$, that $U_{\ell}$ does not store. Then, Algorithm \ref{alg:shuffling_lemma} specifies a protocol where each server in $G$ can multicast a coded packet useful to the other $k-1$ servers such that after $k$ such transmissions each of them can recover its missing chunk. The total number of bits transmitted in this protocol is $Bk/(k-1)$.
\end{lemma}

\begin{algorithm}[!t]
\KwIn{Group of servers $G=\{U_1,\dots,U_k\}$,\\
data chunks $\{\mathcal{D}_{[j]}: U_{j}\in G\}$ s.t. $\forall j, D_{[j]}\in U_l$ where $l\neq j$ and $D_{[j]}\notin U_j$.}
{
\abovedisplayskip=0pt
\belowdisplayskip=0pt
\For{each chunk $\mathcal{D}_{[j]}$}{
Split the chunk into $k-1$ disjoint packets $$\mathcal{C}=\{\mathcal{D}_{[j]}[i]:i=1,\dots,k-1\}.$$\\
Consider a complete bipartite graph with vertex set\\
$\{G\setminus\{U_j\}, \mathcal{C}\}$ and choose a matching $H^{[j]}$\\
within the graph s.t. each node in $G\setminus\{U_{j}\}$ is\\
matched to a node in $\{\mathcal{D}_{[j]}[1],\dots,\mathcal{D}_{[j]}[k-1]\}$.\\
$H^{[j]}(U_l)$ denotes the right neighbor of $U_l$ in $H^{[j]}$.\\
}
\For{each server $U_m\in G$}{
$U_m$ broadcasts\footnotemark\
\begin{equation}
\label{eq:lemma_broadcast}
\Delta_m=\underset{j}{\oplus}H^{[j]}(U_m).
\end{equation}
}
}
\caption{Shuffling algorithm of Lemma \ref{lemma:shufffling_lemma}.}
\label{alg:shuffling_lemma}
\end{algorithm}

\footnotetext{The operation in eq. \eqref{eq:lemma_broadcast} is a bitwise XOR.}

\section{Single-job case}
\subsection{Overview of the method}
The process starts by generating the SPC code as described in Section \ref{sec:primer_resolvable}. The code controls how many subfiles the data set needs to be split into and the corresponding resolvable design gives the assignment of subfiles to servers (for the Map phase). The workers receive the corresponding subfiles from the master node and process them during the Map phase. The resulting intermediate values are encoded into packets by each worker. Specifically, each server computes one encoded packet for each shuffling group it will be participating into during the communication phase. Subsequently, they form groups of fixed size and communicate during the Shuffle phase. In the Shuffle phase, each worker receives intermediate data that it needs in order to perform its reduction operations. These encoded packets are decoded using locally computed intermediate data. Finally, the servers reduce their assigned functions and return all results to the master node.

\subsection{Problem formulation}
\label{sec:formulation_one_job}
We now discuss the problem formulation more formally, based closely on \cite{LiMA16}. In the single-job scenario, the goal is to process one distributed MapReduce job. Let $\mathcal{W}$ denote the data set. There are $N$ input files that correspond to equal-sized and disjoint parts of $\mathcal{W}$. There are $Q$ arbitrary output functions that need to be computed across these $N$ files. There are a total of $K$ servers $U_1,\dots,U_K$. The files will be denoted by $w_1, \dots, w_N$ and the output functions by $\phi_j, j = 1, \dots, Q$. Each function $\phi_j$ depends on all the files $w_1, \dots, w_N$.
We assume that the $j$-th function can be computed by a Map phase followed by a Reduce phase, i.e., $\phi_j(w_1, \dots, w_N) = h_j(g_{j,1}(w_1), \dots, g_{j,N}(w_N))$.
Here, $\bfg_n = (g_{1,n}, \dots, g_{Q,n})$ maps the file $w_n$ into $Q$ intermediate values $\nu_{j,n}, j = 1, \dots, Q$ each of which is assumed to be of size $B$ bits. The function $h_j$ maps the intermediate values $\nu_{j,n}$ on all files into a ``reduced" value $h_j(g_{j,1}(w_1), \dots, g_{j,N}(w_N))$.

\begin{example}
\label{eg:example_1}
Suppose that we consider the problem of computing $Q=4$ functions in a data set consisting of $N=4$ files on a cluster with $K=4$ servers. The files are $w_1, \dots, w_4$ and the functions are $\phi_1, \dots, \phi_4$, e.g., $\phi_1(w_1, \dots, w_4)$ would be the evaluation of $\phi_1$ on the entire data set. Let us assume that the $i$-th server is assigned file $w_i$ for all values of $i$. In the Map phase, server $i$ computes $\bfg_i$ on its assigned file $w_i$ for $i = 1, \dots, 4$. In the Reduce phase, we can see that, e.g., $\phi_1(w_1, \dots, w_N)$ can be computed as $\phi_1(w_1, \dots, w_N) = h_1(g_{1,1}(w_1), \dots, g_{1,N}(w_N))$.
\end{example}

As noted in Section \ref{sec:intro}, there are several MapReduce jobs where the Shuffle phase is rather time-intensive. Thus, when operating on a tradeoff between communication and computation, i.e., one could choose to increase the computation load of the system by processing the same file at $r > 1$ servers. This would in turn reduce the number of intermediate values it needs in the Reduce phase.
For the remainder of the paper, we refer to $r$ as the computation load.

\begin{table}[t]
\centering
\caption{Coded transmissions in all groups of Example \ref{eg:N4_example}.}
\resizebox{0.75\columnwidth}{!}{
\begin{tabular}{ |c|c|c| }
\hline
Group&Server&Transmission\\
\hline
\multirow{3}{*}{$G_1$}
&$U_1$&$p(\nu_{3,2})[1]\oplus p(\nu_{6,1})[2]$\\
&$U_3$&$p(\nu_{6,1})[1]\oplus p(\nu_{1,3})[2]$\\
&$U_6$&$p(\nu_{3,2})[2]\oplus p(\nu_{1,3})[1]$\\
\hline
\multirow{3}{*}{$G_2$}
&$U_1$&$p(\nu_{5,2})[1]\oplus p(\nu_{4,1})[1]$\\
&$U_4$&$p(\nu_{5,2})[2]\oplus p(\nu_{1,4})[1]$\\
&$U_5$&$p(\nu_{4,1})[2]\oplus p(\nu_{1,4})[2]$\\
\hline
\multirow{3}{*}{$G_3$}
&$U_2$&$p(\nu_{5,3})[1]\oplus p(\nu_{3,4})[1]$\\
&$U_3$&$p(\nu_{5,3})[2]\oplus p(\nu_{1,1})[1]$\\
&$U_5$&$p(\nu_{3,4})[2]\oplus p(\nu_{1,1})[2]$\\
\hline
\multirow{3}{*}{$G_4$}
&$U_2$&$p(\nu_{6,4})[1]\oplus p(\nu_{4,3})[1]$\\
&$U_4$&$p(\nu_{6,4})[2]\oplus p(\nu_{2,2})[1]$\\
&$U_6$&$p(\nu_{2,2})[2]\oplus p(\nu_{4,3})[2]$\\
\hline
\end{tabular}
}
\quad
%
\label{table:groups_eg1}
\vspace{-0.2in}
\end{table}

\begin{definition}
\label{def:comm_load}
The communication load $L \in [0,1]$ of a certain single-job scheme is defined as the ratio of the total number of bits transmitted in the data shuffling phase to $QNB$.
\end{definition}

In Example \ref{eg:example_1}, for the baseline approach, at the end of the Map phase, each server needs three values from the other servers. Thus, the total number of bits transmitted would be $4 \times 3 \times B = 12B$. Thus, the communication load of the system will be $L = 12B/16B = 3/4$.

Example \ref{eg:N4_example} that follows examines a single job and demonstrates that increasing $r$ can translate into lower communication loads compared to the baseline method.

\begin{example}
\label{eg:N4_example}
Consider a system with $K=6$ servers, a computation load of $r=3$ (i.e., each Map task will be assigned to 3 distinct servers) and $Q=6$ functions to be computed. Each of these functions depends on the entire data set and will be assigned to one server for the Reduce phase. In our approach we would subdivide the original job into $N=4$ files that will be assigned to the servers as demonstrated in Table \ref{table:placement_eg_N4_example}. 
At the end of the Map step, each server would have computed the $Q$ functions on its assigned Map files. Suppose that the $i$-th server is responsible for reducing the $i$-th function. This would imply, for example, that server $U_1$ needs the first function's evaluation on files $w_3$ and $w_4$.

The key idea of our approach is for each server to transmit a packet that is simultaneously useful to multiple servers. For example, let us consider the group of servers $G_1=\{U_1,U_3,U_6\}$ that were assigned files $\{w_1, w_2\}, \{w_1,w_3\}$ and $\{w_2, w_3\}$, respectively.
At the end of the Map phase, e.g., server $U_1$ wants $\nu_{1,3}$, server $U_3$ wants $\nu_{3,2}$ and server $U_6$ wants $\nu_{6,1}$. We assume that $\nu_{j,n}$ can be encapsulated into a packet with size $B$ bits, denoted by $p(\nu_{j,n})$. Furthermore, assume that this packet can be subdivided into two parts $p(\nu_{j,n})[1]$ and $p(\nu_{j,n})[2]$ (with size $B/2$ bits).


Now consider Table \ref{table:groups_eg1}. Note that server $U_1$ contains files $w_1$ and $w_2$ and can therefore compute all $Q$ functions associated with them. Thus, it can transmit $p(\nu_{3,2})[1] \oplus p(\nu_{6,1})[2]$ as specified in row 1 of the top-left block in Table \ref{table:groups_eg1}. Note that this transmission is {\it simultaneously} useful to both servers $U_3$ and $U_6$. In particular, server $U_3$ already knows $p(\nu_{6,1})[2]$ and can therefore decode $p(\nu_{3,2})[1]$ which it wants. Likewise, server $U_6$ already knows $p(\nu_{3,2})[1]$ and can decode $p(\nu_{6,1})[2]$ that it wants. In a similar manner, it can be verified that each of the transmissions in Table \ref{table:groups_eg1} benefits two servers of the corresponding group. The process of picking the servers to consider together can be made systematic; in addition to server group $G_1$ that we just considered, we can pick three others: $G_2=\{U_1,U_4,U_5\}$, $G_3=\{U_2,U_3,U_5\}$ and $G_4=\{U_2,U_4,U_6\}$ which will result in all the servers obtaining their desired values.

The total number of bits transmitted in this case is therefore $4 \times 3 \times B/2 = 6B$, where $B$ is the size of each intermediate value $\nu_{i,j}$ in number of bits; thus, the communication load is $\frac{6B}{QNB} = 0.25$. In contrast, uncoded transmission from the different servers would have required a total of $2 \times 6 \times B = 12B$ bits to be transmitted, corresponding to a communication load of $0.5$ which is twice of the proposed approach. We emphasize that if $r>1$ an uncoded scheme will also assign multiple copies of each Map task to different servers; all of the servers need to return the values. This assumption is taken into account in our communication load analysis based on Definition \ref{def:comm_load} as it facilitates a fair comparison across different methods and is implemented in all of the algorithms (\emph{cf.} \cite{KKCT} and \cite{USCTS}).

We note here that the authors of \cite{LiMA16} promise a communication load of $L_{coded}(r) = \frac{1}{r} (1 - \frac{r}{K})\approx 0.17$. In general, the possible values of $r$ for that scheme are $\{1, \dots, K\}$. However, crucially this result assumes that $N = \binom{K}{r} \eta_1$, where $\eta_1$ is a positive integer.
It is evident that $N$ grows very rapidly for their scheme. In Section \ref{sec:results_one_job}, we demonstrate that in real-life experiments this idealized analysis is problematic.
\end{example}

We acknowledge that some MapReduce algorithms may be impacted by \emph{data skewness} \cite{Kwon:2012:SMS:2213836.2213840}, a situation when certain Map or Reduce tasks may take significantly longer to process than others. However, TeraSort as well as distributed matrix-vector multiplication (considered in Section \ref{sec:mat_vec_multiplication}) do not suffer from this issue \cite{DBLP:journals/debu/KwonRBH13}. For these problems our assumption of homogeneous mappers and reducers is a reasonable one. This justifies the fact that both prior and proposed methods split the data set into equal-sized subfiles each mapped to an intermediate value of a fixed number of bits. Also, our algorithms deliberately assign roughly equal number of Reduce operations to all workers. We emphasize that the focus of our work is not solving all issues with respect to Shuffle phase traffic reduction in MapReduce systems but to reveal the potential of coding-theoretic methods in this area.
\subsection{From resolvable designs to protocol specification}
\label{subsec:code_protocol}

\begin{algorithm}[!t]
\KwIn{File $\mathcal{W}$, $Q$ functions, number of servers $K = k \times q$. $K$ divides $Q$.
}
{
\abovedisplayskip=0pt
\belowdisplayskip=0pt
Use a $(k,k-1)$ SPC code to generate a design $(\mathcal{X},\mathcal{A})$.\\
Split $\mathcal{W}$ into $q^{k-1}$ disjoint files, $w_{1},\dots,w_{q^{k-1}}$.\\
Assign files to servers such that server $B_{i,j}$ is assigned file $w_\ell$ if $\ell \in B_{i,j}$.\label{alg:file_match}\\
Partition $[Q]$ into $K$ equal parts to obtain the sets $\phi^{B_{i,j}}$ for $i = 1, \dots, k \text{ and } j = 0, \dots, q-1$.
Execute the Map phase on each of the servers.\\
Choose all possible sets $\{B_{1,j_1}, B_{2, j_2}, \dots, B_{k,j_k}\}$ where $j_\ell \in \{0, \dots, q-1\}$,\label{alg:group_rule} such that $\cap_{\ell=1}^k B_{\ell, j_\ell} = \emptyset$ and store them in a collection $\calG$.\\
\For{$\gamma \in [Q/K]$}{
\For{each group $G=\{B_{1,j_1}, B_{2, j_2}, \dots, B_{k,j_k}\}\in \calG$}{
Determine $\mathcal{D}_{[\ell]} = \nu_{\phi^{B_{i,j}}[\gamma], \cap_{k\neq \ell} B_{k,j_k}}$ for $\ell = 1, \dots, k$ used in Algorithm \ref{alg:shuffling_lemma} and execute this algorithm to exchange this data among the servers in $G$.\\
}
}
Execute Reduce phase on each of the servers.
}
\caption{Proposed single-job protocol.}
\label{Alg:protocol}
\end{algorithm}

We assume that $Q$ is a multiple of $K$. In Algorithm \ref{Alg:protocol}, we present the protocol which can be understood as follows. We choose an integer $q$ such that $q$ divides $K$, i.e., $K = k \times q$. Next, we form a $(k,k-1)$ SPC code and the corresponding resolvable design using the procedure in Section \ref{sec:primer_resolvable}. The point set $\calX = [q^{k-1}]$ and the block set $\calA$ will be such that $|\calA| = kq$. The blocks of $\calA$ will be indexed as $B_{i,j}, i = 1, \dots, k$ and $j = 0, 1, \dots, q-1$.

We associate the point set $\calX$ with the files, i.e., $N = |\calX| = q^{k-1}$ and the block set $\calA$ with the servers. For the sake of convenience we will also interchangeably work with servers indexed as $U_{1}, \dots, U_{K}$ with the implicit understanding that each $U_i, i \in [K]$ corresponds to a block from $\mathcal{A}$.  The Map task assignment follows the natural incidence between the points and the blocks, i.e., server $B_{i,j}$ is responsible for executing the Map tasks on the set of files $\text{Map}[B_{i,j}] = \{w_\ell ~|~ \ell \in B_{i,j}\}$. Thus, at the end of the Map phase, server $B_{i,j}$ has computed the $Q$ intermediate values on the files in $\text{Map}[B_{i,j}]$.

Recall that we assume that $K$ divides $Q$. To make load balancing fair we assign $Q/K$ functions to each of the $K$ servers per job for the Reduce phase. This assumes that all Q functions are computed on every file during the Map phase and sent to the appropriate server. However, if $Q$ is a multiple of $K$, then each transmitter can transmit a coded packet in which each term is the concatenation of $Q/K$ intermediate values, one for each function of the receiver. An alternative approach would be to have the servers communicate $Q/K$ times, one for each intermediate value needed by a server (this idea is used in Algorithm \ref{Alg:protocol}). We let $\phi^{B_{i,j}} \subset [Q]$ represent the set of functions assigned for reduction to server $B_{i,j}$. The sets $\phi^{B_{i,j}}$ form a partition of $[Q]$. For ease of notation, we let $\phi^{B_{i,j}}[\ell]$ represent the $\ell$-th function in the set $\phi^{B_{i,j}}$; $\ell\in[Q/K]$.

Following the Map phase, in the Shuffle phase, each server $B_{i,j}$ needs intermediate values from other servers so that it has enough information to reduce the functions in $\phi^{B_{i,j}}$. In this step we transmit coded packets that are simultaneously useful to multiple servers. Towards this end we form a collection of server groups by choosing one block from each parallel class according to the rule in Step \ref{alg:group_rule} of the protocol, i.e., we choose servers $B_{1,j_1}, B_{2, j_2}, \dots, B_{k,j_k}$ such that $\cap_{\ell=1}^k B_{\ell, j_\ell} = \emptyset$. For a given server group $G$ (of size $k$) we utilize Algorithm \ref{alg:shuffling_lemma}.

\begin{table*}[!t]
\centering
\caption{Measurements for sorting 12GB data on 16 server nodes without coding.}
\resizebox{1.2\columnwidth}{!}{
\begin{tabular}{ |P{0.7cm}|P{0.7cm}|P{1cm}|P{1cm}|P{1cm}|P{1.5cm}|P{1cm}| }
 \hline
 Map&Pack&Shuffle&Unpack&Reduce&Total Time&Rate\\
 (sec.)&(sec.)&(sec.)&(sec.)&(sec.)&(sec.)&(Mbps.)\\
 \hline
 3.36&2.55&999.84&2.19&12.45&1020.39&100.80\\
 \hline
\end{tabular}
}
\label{table:uncoded}
\end{table*}

\begin{table*}[!t]
\centering
\caption{MapReduce time for sorting 12GB data on 16 server nodes including the memory allocation cost.}
\label{table:tests}
\resizebox{2\columnwidth}{!}{
\begin{tabular}{ |P{2.2cm}||P{1.1cm}|P{0.7cm}|P{1cm}|P{1.1cm}|P{1.1cm}|P{0.9cm}|P{1cm}|P{1cm}|P{0.9cm}|P{0.9cm}|P{0.9cm}|P{0.7cm}| }
\hline
&\multirow{3}{*}{CodeGen}&\multirow{3}{*}{Map}&\multirow{3}{*}{\makecell{Pack/\\ Encode}}&\multirow{3}{*}{Shuffle}&\multirow{3}{*}{\makecell{Unpack/\\ Decode}}&\multirow{3}{*}{Reduce}&\multicolumn{2}{c|}{Total Time}&\multicolumn{2}{c|}{\multirow{2}{*}{Speedup}}&\multirow{3}{*}{Rate}&\multirow{4}{*}{N}\\
&\multirow{3}{*}{(sec.)}&\multirow{3}{*}{(sec.)}&\multirow{4}{*}{(sec.)}&\multirow{3}{*}{(sec.)}&\multirow{4}{*}{(sec.)}&\multirow{3}{*}{(sec.)}&\multicolumn{2}{c|}{(sec.)}&\multicolumn{2}{c|}{}&\multirow{3}{*}{(Mbps.)}&\\
\cline{8-9}\cline{10-11}
&&&&&&&\multirow{2}{*}{w/MA}&w/out MA&\multirow{2}{*}{w/MA}&w/out MA&&\\
\hline
Uncoded: &-&5.71&11.75&1105.64&4.46&12.88&1140.44&1126.68&&&100.83&16\\
Prior: $r=3$&5.82&17.94&229.80&455.05&6.23&14.54&729.38&496.76&$1.56\times$&$2.27\times$&64.79&560\\
Prior: $r=5$&26.78&29.99&1000.15&297.28&8.16&16.47&1378.83&490.28&$0.83\times$&$2.30\times$&61.04&4368\\
Prior: $r=8$&38.41&51.03&1128.16&-&-&-&-&-&-&-&-&12870\\
Proposed: $r=4$&0.64&25.91&9.93&307.15&6.91&17.29&367.83&352.91&$3.10\times$&$3.19\times$&88.47&64\\
Proposed: $r=8$&0.61&62.46&26.22&127.43&8.38&17.85&242.95&204.58&$4.69\times$&$5.51\times$&62.68&128\\
\hline
\end{tabular}
}
\end{table*}

\subsection{Proof of correctness and communication load analysis}

We now prove that the proposed protocol allows each server $B_{i,j}$ to recover enough information at the end of the Shuffle phase. As the protocol is symmetric with respect to blocks, we equivalently show that server $B_{1,j_1}$ is satisfied. Note that $|B_{1,j_1}| = q^{k-2}$. For the purposes of our arguments below, we assume that $Q = K$. The case when $Q$ is an integer multiple of $K$ is quite similar. In this case, with some abuse of notation, since $\phi^{B_{1,j_1}}$ is a singleton set, we use $\phi^{B_{1,j_1}}$ to actually represent the function index itself. It is therefore clear that $B_{1,j_1}$ needs the intermediate values $\nu_{\phi^{B_{1,j_1}},n}$ for $n \in [q^{k-1}] \setminus B_{1,j_1}$.

Now consider the construction of the server groups. Let $G$ be a server group where $B_{1,j_1}$ is chosen from $\calP_1$, i.e., $G = \{B_{1,j_1},B_{2,j_2}, \dots, B_{k,j_k}\}$. The following lemma (proved in \cite{TangR18}), shows that the intersection of {\it any} $k-1$ blocks from $k-1$ distinct parallel classes is always of size $1$.

\begin{lemma}
\label{lemma:intersect}
Consider a proposed resolvable design $(X,\mathcal{A})$ constructed with parameters $k$ and $q$ and parallel classes $\mathcal{P}_1,\dots,\mathcal{P}_k$. If we pick $k-1$ blocks $B_{i_1,l_1},\dots,B_{i_{k-1},l_{k-1}}$ (where $i_j\in [k]$, $l_j\in\{0,\dots,q-1\}$) from distinct parallel classes $\mathcal{P}_{i_1},\dots,\mathcal{P}_{i_{k-1}}$, then $|\cap_{j=1}^{k-1}B_{i_j,l_j}|=1$.
\end{lemma}

Furthermore, note that the intersection of all the blocks in $G$ is empty (\emph{cf.} Step \ref{alg:group_rule} of Algorithm \ref{Alg:protocol}). There is one-to-one correspondence between this setup and Lemma \ref{lemma:shufffling_lemma}. The group of servers on which we will apply the lemma is precisely $G$. Also, observe that server $B_{\ell,j_\ell}$ misses the unique file $\cap_{k \neq \ell} B_{k,j_k}$ that all other servers in $G$ share (\emph{cf.} Lemma \ref{lemma:intersect}) and $B_{\ell,j_\ell}$ will be reducing the function $\phi^{B_{\ell,j_\ell}}$ (note that we have dropped the index $\gamma$ from the intermediate value corresponding to $\mathcal{D}_{[\ell]}$ from Algorithm \ref{Alg:protocol} due to our assumption that $Q=K$). Hence, the correspondence of  intermediate values to chunks of Lemma \ref{lemma:shufffling_lemma} is $\mathcal{D}_{[\ell]} = \nu_{\phi^{B_{\ell,j_\ell}}, \cap_{k \neq \ell} B_{k,j_k}}$.

We conclude the proof by observing that a given block, e.g., $B_{\ell,j_\ell}$ participates in $q^{k-2} (q-1) = q^{k-1} - q^{k-2}$ server groups each of which allow it to obtain distinct intermediate values. This can be seen as follows. Suppose for instance, that $ \cap_{k \neq \ell} B_{k,j_k} = \cap_{k \neq \ell} B_{k,j_k'}$ where $j_m \neq j_m'$ for at least one value of $m \in [k] \setminus \{\ell\}$. In this case, we note that the equality above implies that
$\cap_{k \neq \ell} B_{k,j_k} \bigcap \cap_{k \neq \ell} B_{k,j_k'} \neq \emptyset$. This is a contradiction, because $B_{m,j_m} \cap B_{m,j_m'} = \emptyset$ as they are two blocks belonging to the same parallel class.

Therefore, since $B_{\ell,j_\ell}$  is missing exactly $q^{k-1} - q^{k-2}$ intermediate values, it follows that at the end of the Shuffle phase it is satisfied. By symmetry,  all servers are satisfied.

Next, we present the analysis of the communication load of our algorithm. In the uncoded case, each server needs $QN/K$ intermediate values $\nu_{j,n}$'s to execute its Reduce phase. Note that each server already has $rN/K \times Q/K$ of them. Thus, the communication load is given by
\begin{align*}
L_{\mathrm{uncoded}}^{\mathrm{single}} = \frac{K(QN/K - rQN/K^2)B}{QNB}
= 1 - \frac{r}{K}.
\end{align*}
On the other hand, for our scheme, the number of bits transmitted in the Shuffle phase is given by
$$ q^{k-1}(q-1) \cdot B\frac{k}{k-1} \cdot \frac{Q}{K}.$$
Thus, the communication load is given by
\begin{align*}
L_{\mathrm{proposed}}^{\mathrm{single}} = \frac{q^{k-1}(q-1) \cdot B\frac{k}{k-1} \cdot \frac{Q}{K}}{QNB} = \frac{1}{k-1} \bigg{(} 1 - \frac{k}{K} \bigg{)},
\end{align*}
where the second equality above is obtained by using the fact that $N = q^{k-1}$ and $K = kq$.

Next, note that for our proposed scheme the computation load is $k$, i.e., $r=k$. Thus, we reduce the overall communication load by a factor of $\frac{1}{r-1}$ with respect to an uncoded system.
In contrast, the approach in \cite{LiMA16}, reduces the communication load by a factor of $\frac{1}{r}$. However, this comes at the expense of a large $N$ as discussed previously.

\subsection{TeraSort Experimental Results and Discussion}
\label{sec:results_one_job}
We implemented our technique on Amazon EC2 and performed comparisons with the method of \cite{songze_terasort} using their posted software at \cite{USCTS}.
Table \ref{table:uncoded} corresponds to a uncoded TeraSort with $r=1$. It shows that the Shuffle phase which takes $999.84$ seconds, dominates the overall execution time by far. A detailed description of the setup appears in the Appendix.

\begin{table}[!t]
\centering
\caption{TeraSort memory allocation cost percentage.}
\begin{tabular}{ |P{2cm}||P{3.3cm}| }
\hline
& Memory allocation time (\%)\\
\hline
Uncoded: &1.2\\
Prior: $r=3$&31.9\\
Prior: $r=5$&64.4\\
Prior: $r=8$&-\\
Proposed: $r=4$&4.1\\
Proposed: $r=8$&15.8\\
\hline
\end{tabular}
\label{table:memory_percentage}
\end{table}

Table \ref{table:tests} contains the results of TeraSort using our approach and comparisons with the approach in \cite{songze_terasort}. Nearly $130\times10^6$ KV pairs should be sorted. The time required for each phase has been reported. For the total time taken, we have reported the numbers including and excluding the memory allocation time-cost. This is because the results in \cite{songze_terasort} are generated using the code in \cite{USCTS} which explicitly ignores the memory allocation time ({\it cf.} communication with the first author of \cite{songze_terasort}). However, we have observed that for data sets at this scale, dynamic memory allocation on the heap (using the C++ $\texttt{new}$ operator) has a non-negligible impact on the total time. Thus, in our implementation (available at \cite{KKCT}), we measure the memory allocation time as well and we report its fraction on Table \ref{table:memory_percentage}. We emphasize however, that the results in Table \ref{table:tests} indicate that our approach is consistently superior whether or not one takes into account the memory allocation time.


To understand the effect of choosing different values of $N$ (\emph{cf.} Table \ref{table:tests}), we applied our algorithm with different values of $(k,q)=(r,q)$ pairs. 
We observe from Table \ref{table:tests} that if we account for the memory allocation cost, our scheme achieves up to $4.69\times$ speedup compared to the uncoded TeraSort whereas if we ignore this cost our schemes demonstrates an improvement of up to $5.51\times$. Moreover, the gain over the prior coded TeraSort scheme, if we compare the best time reported by each scheme, can go up to $4.69/1.56\approx3.01\times$ (when including memory allocation time) or $5.51/2.30\approx2.4\times$ (when excluding memory allocation time). We note here that the Shuffle phase results corresponding to $r=8$ for prior work could not be obtained as their program crashed. The following inferences can be drawn from Table \ref{table:tests}.

\begin{itemize}
\item The algorithm starts with the CodeGen phase during which all workers generate the resolvable design
based on our choice of the parameters $q$ and $k$. Based on the design, all groups of workers that will be communicating in the Shuffle phase are determined. Next, the data set is split into $N$ files by
the master and the appropriate files are transmitted to each worker.
In our experiments this phase is quite efficient since the number of groups we need to generate and consequently the number of shuffling sub-groups we need to split the group containing all the servers into, is much smaller than that of the prior scheme. For example, let us look at the CodeGen time for $r=3$ of the prior scheme which is $t_1=5.82$. The corresponding number of groups is $g_1={{K}\choose{r+1}}={{16}\choose{4}}=1820$. For our scheme, that time is $t_2=0.64$ and the number of multicast groups is $g_2=q^{r-1}(q-1)=4^3\times3=192$. Now if we try to interpolate our code generation cost from $t_1$, based on our analysis, we would get:
$$t_2^{'}=\frac{g_2}{g_1}\times t_1=\frac{192}{1820}\times5.82\approx 0.61\approx t_2.$$
\item The Map time mainly depends on the computation load $r$. Since $r$ is the number of times the whole data set is replicated and processed across the cluster we expect the Map cost of both coded schemes to be approximately $r$ times higher than that of the uncoded implementation. Indeed, if we look at our scheme for $r=4$ we see that $\frac{25.91}{5.71}\approx4.54$ is a good approximation to $r$.
\item The encoding time of the coded schemes (which is the time it takes so that all servers form the encoded packets that they will be transmitting afterwards) is not directly comparable to the packing of the uncoded approach which stores each intermediate value serially in a continuous memory array to ensure that a single TCP connection is initiated for each intermediate value. Further examination of the internals of C++ dynamic memory allocation (which we used) is beyond the scope of our analysis but one point we emphasize is that we have a significant benefit over the prior scheme during encoding. For $r=8$, we obtain a speedup of $\frac{1128.16}{26.22}\approx43.03$.
This is explained by the fact that in the previous scheme each server participates into much more groups and thus it needs to store more encoded data into its memory.
\item The Shuffle phase is where we can see the advantage of our implementation. For example, when $r=8$, our predicted load will be $1/14$, while the load of the uncoded $r=1$ scheme will be $15/16$. Thus, with the same transmission rate we expect our Shuffle phase to be $13.125$ times faster. However, our obtained transmission rate is approximately $62.68$ Mbps. Thus, the overall gain is expected to be around $8.16$ times. In the actual measurements our gain is $\frac{1105.64}{127.43}\approx8.68$ which is quite close to the prediction. On the other hand, let us consider the prior scheme when $r=5$. In this case the load analysis predicts a gain of $6.82$ assuming that the transmission rates are the same and a gain of $4.13$ when accounting for the different rates. However, the actual gain is $\frac{1105.64}{297.28}\approx 3.72$. Some of these discrepancies can be explained by the fact that the cost of multicasting a message from a server to $n$ receivers is not necessarily $n$ times cheaper than unicasting that message separately to each of the $n$ receivers. 
In particular, in Open MPI there are seven modes of broadcasting a common message to multiple receivers. These include \emph{basic linear}, \emph{chain} and \emph{binary tree} among others. For instance, in a typical binary tree the sender is the root of the tree and the receivers are the descendants of it. The transmission starts from the root and propagates downward. The depth of the tree is logarithmic in the number of nodes so it can achieve a logarithmic speed-up as compared to unicast. Since the details of these implementations fall beyond the scope of our research we have chosen to use the automatic module which selects the transmission algorithm on-the-fly depending on the communicator and message sizes. However, our load analysis corresponds to a \emph{basic linear} broadcast (the sender sends a common message to all receivers one at a time without parallel communication). Hence, our definition of communication load, defined as the total number of bits transmitted divided by the time, provides a theoretical worst-case of the load one could achieve; it also sets a common metric which helps us compare with uncoded approaches and other coded schemes in equal terms. Indeed, for small communicators of size $k$ like in our experiments the MPI quite likely resorts to the basic linear broadcast and the transmitter sends the common packet sequentially to all receivers \cite{Hoefler_MPI_Bcast}. If MPI resorts to a parallel broadcast algorithm such that of a binary tree it won't generally perform all transmissions of a tree level in parallel. However, it will prioritize them such that one child of each transmitter is serviced first while the other is waiting and it will maximize the bandwidth of the connections of the transmitter-receiver pairs which are serviced first.
The overhead of setting the connections is much lesser in our protocol due to the reduced number of groups. Specifically, the \emph{latency} (number of transmissions) of our method is $k$ transmissions per group for a total of $q^{k-1}(q-1)k$ transmissions. Moreover, the lower layers of network protocols introduce additional headers into packets likely to affect more the prior scheme due to smaller payloads.
\end{itemize}

\begin{figure}[t]
\centering
\includegraphics[scale=0.65]{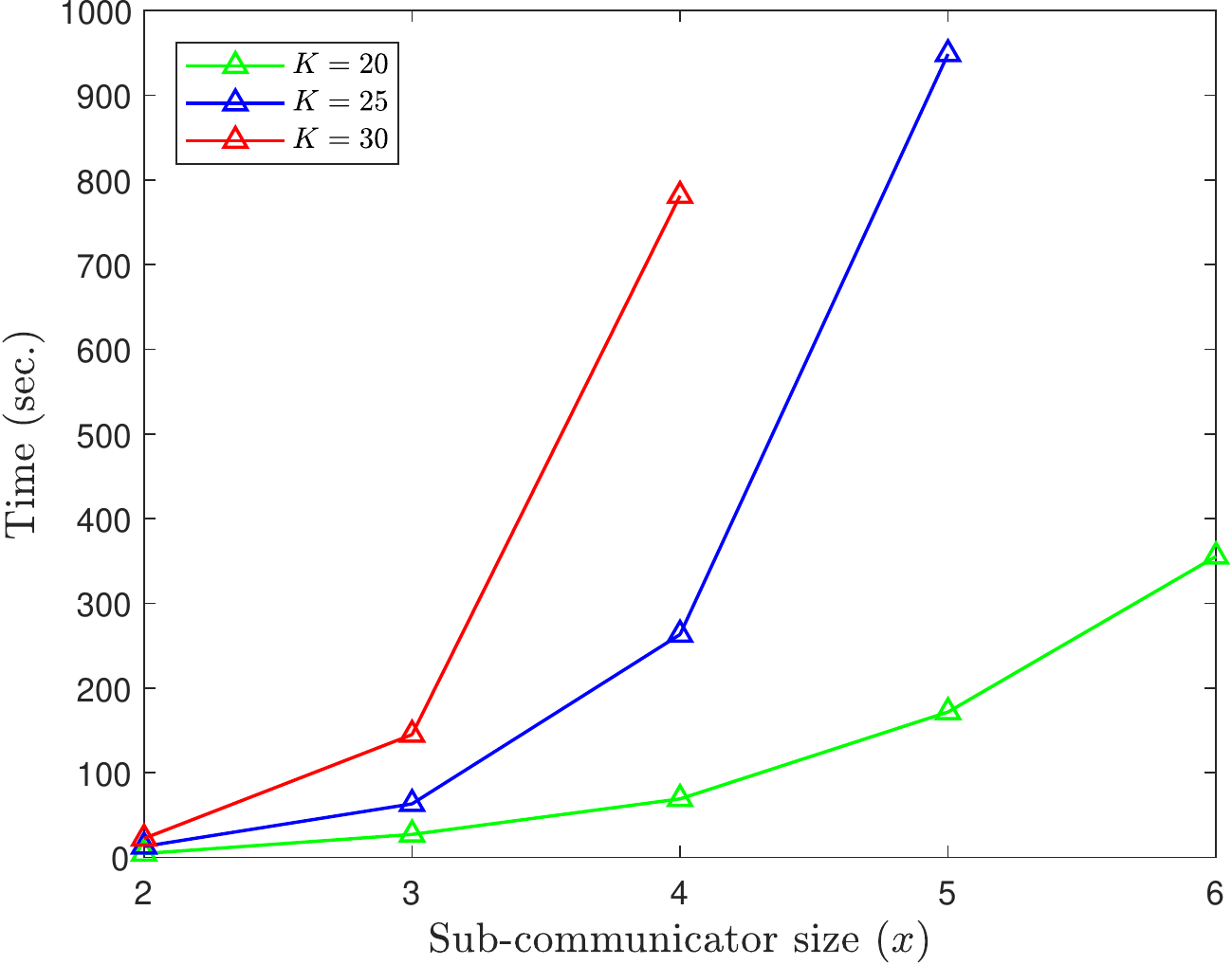}
\caption{\texttt{MPI\_Comm\_Split} execution time.}
\label{fig:comm_split}
\vspace{-0.2in}
\end{figure}

The major issue of the prior TeraSort scheme in \cite{songze_terasort} is the large value of $N$ that it needs. This translates into a large number ($\binom{K}{r+1}$) of server groups in the shuffling phase. This number can be prohibitive for High Performance Computing (HPC) communication protocols like the Message Passing Interface (MPI). This is because all MPI communication is associated with a \emph{communicator} that describes the communication context and an associated group of processes. But, the cost of splitting the initial communicator is non-negligible \cite{Sack2010}. In the case of coded TeraSort of \cite{songze_terasort} the overall communicator needs to be split into ${K}\choose{r+1}$ \emph{intra-communicators} each facilitating the communication within a group. 

We demonstrate the impact of this issue by explicitly measuring the time needed to split the initial communicator of $K$ servers into ${K}\choose{x}$ intra-communicators, each of size $x$ for different values of $K$ and $x$. Let us refer to Fig. \ref{fig:comm_split}.
We see that MPI\_Comm\_Split incurs an exponential cost that can easily dominate the overall MapReduce execution. This clearly indicates that even though the communication load may reduce with increasing $r$ in the scheme of \cite{songze_terasort}, the overall execution time may be adversely affected (see \cite{Sack2010} for more details).

Another point to consider is that the MPI library might support a limited number of communicators. Some indicative examples are those of Open MPI which supports up to $2^{30}-1$ communicators, MPI over InfiniBand, Omni-Path, Ethernet/iWARP and RoCE (MVAPICH) which allows for up to 2000 communicators and High-Performance Portable MPI (MPICH) that limits this number to 16000. Thus, if we have ($K=50$, $r=10$) the number of required groups will be ${50}\choose{11}$ which would exceed these limits. In our method, we could choose $(q,k)=(5,10)$ or $(q,k)=(2,25)$ both of which are below Open MPI communicator limits, requiring 7812500 and 16777216 groups, respectively.

Our experiments indicate that the time consumed in memory allocation can be non-negligible and this is a major issue. We emphasize though that our gains over prior methods hold even if we do not take the memory allocation time into account.

Another interesting aspect of our experiments is that the observed transmission rate appears to change based on the value of $r$. In our experiments we capped the transmission rate at $100$ Mbps. However, the observed rate can be as low as $61.04$ Mbps. As our experiments run on Amazon EC2, we do not have a clear explanation on the underlying reasons. Nevertheless, we point out the rates for our proposed $r=8$ and the prior scheme $r=5$ are quite close.

\section{Multi-Job Case for functions amenable to aggregation}
\label{sec:formulation_camr}


In this section we discuss how resolvable designs can help with processing multiple jobs on a cluster where the underlying functions are amenable to aggregation. Our goal is to process $J$ distributed computing jobs (denoted $\mathcal{J}_1,\dots,\mathcal{J}_J$) in parallel on a cluster with $K$ servers. The data set of each job is partitioned into $N$ disjoint and equal-sized files. The files of the $j$-th job are denoted by $n^{(j)},n=1,\dots,N$. A total of $Q$ output functions, denoted $\phi_q^{(j)},q=1,...,Q$, need to be computed for each job.
Note that these $Q$  functions may be different across different jobs. We examine a special class of functions that possess the \emph{aggregation} property.

\begin{definition}
In database systems, an \emph{aggregate function} $\phi$ is one that is both associative and commutative.
\end{definition}

For example, in jobs with ``linear" aggregation the evaluation of each output function can be decomposed as the sum of $N$ intermediate values, one for each file, i.e., for $q=1,\dots,Q$,
$$\phi_q^{(j)}(1^{(j)},\dots,N^{(j)})=\nu_{q,1}^{(j)}+\nu_{q,2}^{(j)}+\dots+\nu_{q,N}^{(j)},$$
where $\nu_{q,n}^{(j)}=\phi_q^{(j)}(n^{(j)})$ and each such value is assumed to be of size $B$ bits. In what follows we use $\alpha(\nu_{q,1}^{(j)},\dots,\nu_{q,m}^{(j)})$ to denote the aggregation of $m$ intermediate values $\nu_{q,1}^{(j)},\dots,\nu_{q,m}^{(j)}$ of the same function $\phi_q^{(j)}$ and job $\mathcal{J}_j$ into a single compressed value. We assume that it is also of size $B$ bits.


As before, a master machine places the files on the servers according to certain rules. Note that each file is placed on at least one server before initiating the algorithm.

\begin{definition}
The storage fraction $\mu\in[1/K,1]$ of a distributed computation scheme is the fraction of the data set across all jobs that each server locally caches.
\end{definition}

Once again, we assume that $Q$ is divisible by $K$. As we have already discussed, our scheme is easily adapted to that case, we choose to keep the discussion simple and focus on the $Q=K$ case, i.e., each server is reducing one function.
%

The framework starts with the Map phase during which the servers (in parallel)
map every file $n^{(j)}$ that they contain to the values $\{\nu_{1,n}^{(j)},\dots,\nu_{Q,n}^{(j)}\}$.
Following this, the servers multicast the computed intermediate values amongst one another via a shared link in the Shuffle phase.
In the final Reduce phase, server $U_k$ computes (or reduces) $\phi_k^{(j)}(\nu_{k,1}^{(j)},\dots,\nu_{k,N}^{(j)})$ for $j = 1, \dots, J$ as it has all the relevant intermediate values required for performing this operation.

\begin{definition}
The communication load $L$ of a scheme executing $J$ jobs is the total amount of data (in bits) transmitted by the servers during the Shuffle phase normalized by $JQB$.
\end{definition}



Our proposed algorithm will be abbreviated as \emph{CAMR} (Coded Aggregated MapReduce). Our main idea is to again use resolvable designs. However, the interpretation of the design, i.e., the correspondence of the points and blocks with the MapReduce setup is significantly different.

\subsection{Job assignment and file placement}
\label{sec:compressed_job_file_placement}
Our cluster consists of $K$ servers and we choose appropriate integers $q, k$ so that $K=k\times q$. The number of files $N$ needs to be divisible by $k$; we discuss its choice shortly. Next, we form a $(k, k-1)$ SPC code and the resolvable design, as described in Section \ref{sec:primer_resolvable}. The jobs to be executed are associated with the point set $\mathcal{X}=[q^{k-1}]$ so that $J=q^{k-1}$. The block set $\mathcal{A}$ is associated with the servers, i.e, each server corresponds to a  block $B_{i,j}, i=1,\dots,k,$ and $j=0,1,\dots,q-1$.

Job $\mathcal{J}_j$ is processed by (or ``owned" by) the server indexed by $B_{i,l}$ if $j \in B_{i,l}$. Let us denote the owners of $\mathcal{J}_j$ by $X^{(j)} \subset \{U_1, \dots, U_K\}$.
For each job, the data set is split into $k$ \emph{batches} and each batch is made up of $\gamma$ files, for any positive integer $\gamma>1$ (recall that $k|N$); even though there are no other constraints on $\gamma$, it gives us a finer control over the subpacketization level that we want depending on the data set size. The file placement policy is illustrated in Algorithm \ref{alg:placement}.

Each server is the owner of $q^{k-2}$ jobs (block size). For each such job it participates in $k-1$ batches of size $\gamma$, as explained in Algorithm \ref{alg:placement}. Hence, our required storage fraction is
$$\mu=\frac{q^{k-2}\cdot(k-1)\cdot\gamma}{Jk\gamma}=\frac{k-1}{K}.$$

\begin{algorithm}[!t]
\KwIn{$J$ jobs, owner sets $\{X^{(j)}, j=1,\dots,J\}$, $k$ used in SPC code, batch size $\gamma$.}
Set $N=k\gamma$.\\
\For{each job $\mathcal{J}_j$}{
Split the data set of $\mathcal{J}_j$ into $N$ disjoint files $\{1^{(j)},\dots,N^{(j)}\}$ and partition them into\\
\nonl $k$ batches of $\gamma$ files each.\\
Let $X^{(j)}=\{U_{i_1},\dots,U_{i_k}\}$. Label each batch with a distinct index of an owner so that\\
the batches are
$\mathcal{B}=\{\mathcal{B}_{[i_1]}^{(j)},\dots,\mathcal{B}_{[i_k]}^{(j)}\}$.\\
\For{each owner $U_{k'}\in X^{(j)}$}{
Store all batches in $\mathcal{B}$ except $\mathcal{B}_{[k']}^{(j)}$ in server $U_{k'}$.
}
}
\caption{File placement.}
\label{alg:placement}
\end{algorithm}

\subsection{Map phase}
During this phase, each server maps all the files of each job it has partially stored, for all output functions. The resulting intermediate values have the form $\nu_{q,n}^{(j)}=\phi_q^{(j)}(n^{(j)}),\quad q\in[Q],\ n\in[N],\ j\in[J]$.

At the end of the Map phase, for each job $\mathcal{J}_j$, each mapper combines all those values $\nu_{q,n}^{(j)}$ that are indexed with  the same $q$ and $j$ (in other words, associated with the same function and job) and belong to the same batch of files; we have already referred to this operation as aggregation.

\begin{figure*}[t]
\centering
\includegraphics[scale=0.17]{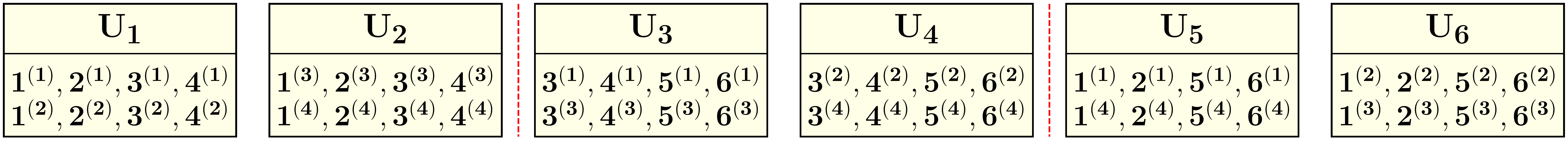}
\caption{Proposed placement scheme for $K=6$ servers and $N=6$ files per computing job for $J=4$ jobs. The dotted lines show the partition of the servers into parallel classes.}
\label{fig:placement_ex1}
\vspace{-0.2in}
\end{figure*}

\subsection{Shuffle phase}
\label{sec:shuffle_scheme}
The CAMR scheme carries out the data shuffling phase in three stages. The first two stages use Algorithm \ref{alg:shuffling_lemma} of Lemma \ref{lemma:shufffling_lemma} introduced for the single-job case.

We will be focusing on a server $U_{k'}$, associated with a block, say $B_{x,y}$, and we will argue that $U_{k'}$ is able to recover all missing aggregate values at the end of the Shuffle phase. Based on Algorithm \ref{alg:placement}, $U_{k'}$ stores batches $\mathcal{B}_{[z]}^{(j)}$ for all values of $(j,z)$ s.t. $j\in B_{x,y}$ and $z\neq k'$; those are the bathes $U_{k'}$ stores for all the jobs it owns. But $U_{k'}$ misses one batch for each of these jobs which is $\mathcal{B}_{[k']}^{(j)}$ for all values of $j$ s.t. $j\in B_{x,y}$; in addition, $U_{k'}$ does not store any batches of the remaining jobs, i.e., it misses the batches $\mathcal{B}_{[z]}^{(j)}$ for all values of $(j,z)$ s.t. $j\notin B_{x,y}$.

\begin{enumerate}[wide, labelwidth=!, labelindent=0pt]
\item \textbf{Stage 1}: In this stage, only owners of each job communicate among themselves. Let us fix a job $\mathcal{J}_j$ that $U_{k'}$ owns and consider the servers in $X^{(j)}\setminus \{U_{k'}\}$ of cardinality $k-1$ (\emph{cf.} Algorithm \ref{alg:placement}). During the Map phase, each server in that subset has computed an aggregate needed by the remaining owner $U_{k'}$ which is (note from Algorithm \ref{alg:placement} that batch $\mathcal{B}_{[k']}^{(j)}$ is not available in $U_{k'}$)
$$\alpha_{[k']}^{(j)}=\alpha(\{\nu_{k',n}^{(j)}:\ n\in \mathcal{B}_{[k']}^{(j)}\}).$$

Let us keep the job $\mathcal{J}_j$ fixed. Then, if we repeat the above procedure for all owners $U_p\in X^{(j)}$ we can identify the aggregates $\alpha_{[p]}^{(j)}$. Each of these values is needed by exactly one owner $U_{p}$.

There is an immediate correspondence between this setup and Lemma \ref{lemma:shufffling_lemma} which is $G=X^{(j)}\quad \text{and}\quad \mathcal{D}_{[p]}=\alpha_{[p]}^{(j)}$ for $j=1,\dots,J$. Hence, Algorithm \ref{alg:shuffling_lemma} can be utilized here so that each owner of $\mathcal{J}_j$, after receiving $k-1$ such values (one from every other owner of that particular job), can decode all of its missing aggregates for job $\mathcal{J}_j$.
We can repeat this process for every value of $j$, i.e., for every job. In total, $J$ groups of servers (the owner set of each job), each of size $k$ will be communicating among themselves in this stage.

At the end of stage 1, worker $U_{k'}$ (block $B_{x,y}$) should have recovered all needed intermediate values of batches of the form $\mathcal{B}_{[k']}^{(j)}$ for all values of $j$ s.t. $j\in B_{x,y}$.

\item \textbf{Stage 2}: In this stage, we form communication groups of both owners and non-owners of a job, so that the latter can recover appropriate data to reduce their functions.

Towards this end, we form collections of server groups by choosing one block from each parallel class based on a simple rule. We choose a group of servers $G = \{B_{1,j_1}, B_{2, j_2}, \dots, B_{k,j_k}\}$ such that $\cap_{\ell=1}^k B_{\ell, j_\ell} = \emptyset$. Without loss of generality, assume that $U_{k'}\in G$.
If we remove $U_{k'}$ from $G$, the servers in the corresponding subset $P=G\setminus \{U_{k'}\}$ of cardinality $|P|=k-1$ jointly own a job, say $\mathcal{J}_j$, that the remaining server $U_{k'}$ does not (\emph{cf.} Lemma \ref{lemma:intersect}). In addition, based on the aforementioned file placement policy (\emph{cf.} Algorithm \ref{alg:placement}), they share the batch of files $\mathcal{B}_{[l]}^{(j)}$ for that common job and some $U_l\in X^{(j)}$. Note that $U_l$ does not contain the batch $\mathcal{B}_{[l]}^{(j)}$.

The following simple observation is important.
\begin{observation}
By construction, $U_l$ is precisely the remaining owner of $\mathcal{J}_j$ and it should lie in the parallel class that none of the other owners belong to. This is precisely the same class in which $U_{k'}$ lies.
\end{observation}

During the Map phase, each server in $P$ has computed an aggregate needed by $U_{k'}$ which is
\begin{equation}
\label{eq:aggreagates_stage2}
\beta_{[k']}^{(j)}=\alpha(\{\nu_{k',n}^{(j)}:\ n\in \mathcal{B}_{[l]}^{(j)}\}).
\end{equation}

As in stage 1, Lemma \ref{lemma:shufffling_lemma} fits in this description and Algorithm \ref{alg:shuffling_lemma} defines the communication scheme; the shuffling group is $G$ and each server $U_{p}\in G$ needs its missing chunk $\mathcal{D}_{[p]}=\beta_{[p]}^{(j)}$ for the unique batch that all servers in $P$ share.

Server $U_{k'}$ participates in $q^{k-2}(q-1)$ such groups $G$ satisfying the aforementioned rule. For each such $G$, $U_{k'}$ does not own a job (and a corresponding batch) that the servers in $G\setminus\{U_{k'}\}$ own. The missing batch is exactly $\mathcal{B}_{[l]}^{(j)}$ for some $l$ such that $U_l$ lies in the same parallel class as $U_{k'}$. At the end of stage 2, $U_{k'}$ is able to decode $q^{k-2}(q-1)$ aggregates of the form in eq. \eqref{eq:aggreagates_stage2} one for each job it does not own.

%
%

\item \textbf{Stage 3}: Each server is still missing values for jobs that it is not owner of from stage 2. Now, servers communicate within parallel classes. We emphasize the following observation.

\begin{observation}
All values that server $U_{k'}$ still needs can be aggregated and transmitted by a single owner-server in the same parallel class that $U_{k'}$ belongs to. This server is unique and transmits one aggregate value of its jobs to every other server in the same parallel class.
\end{observation}

The proof of the above observation follows from stage 2 and by the resolvability property of our design. Let us fix a shuffling group in stage 2, say $G$, a subset $P=G\setminus \{U_{k'}\}$ and focus on the excluded server $U_{k'}$. The servers in $P$ jointly own a unique job $\mathcal{J}_j$ that $U_{k'}$ misses. The remaining owner of $\mathcal{J}_j$ is some $U_l$ that lies in the same parallel class as $U_{k'}$. Note that stage 2 has already allowed us to recover the aggregate on the unique batch of $\mathcal{J}_j$ that the servers in $P$ share; this batch is not contained in $U_l$. However, based on Algorithm \ref{alg:placement}, $U_l$ contains all the other batches associated with $\mathcal{J}_j$ and can hence compute the aggregate function on them. This is exactly what happens in stage 3 for each server.


More formally, we have the following argument. Recall that the $i$-th class is $\mathcal{P}_i=\{B_{i,j},j=0,\dots,q-1\}$ and fix a job $\mathcal{J}_j$ that a server $U_k\in \mathcal{P}_i$ owns and $U_{k'}\in \mathcal{P}_i$ does not. Then $U_k$ transmits
\begin{equation}
\label{eq:shuffle_stage3}
\Delta_k^{\text{stage 3}}=\alpha\Bigg(\bigcup\limits_{l:U_l\in X^{(j)}\setminus \{U_k\}}\{\nu_{k',n}^{(j)}:\ n\in \mathcal{B}_{[l]}^{(j)}\}\Bigg)
\end{equation}
to $U_{k'}\in\mathcal{P}_i$; obviously, $U_{k'}\notin X^{(j)}$. We will do this process for every job that $U_k$ owns and $U_{k'}$ does not. Finally, we will take every pair $(U_{k'},U_k)$ of servers in that parallel class $\mathcal{P}_i$ and repeat the procedure for all parallel classes.

By the end of this last stage 3, $U_{k'}$ has received all missing values that it needs for the Reduce phase. Since the above analysis holds for any value of $k'$, we have shown that our communication scheme serves its purpose and all workers have the necessary data to reduce their functions.

%
\end{enumerate}


\subsection{Reduce phase}
Using the values it has computed and received, $U_k$ reduces $\phi_k^{(j)}(1^{(j)},\dots,N^{(j)})=\alpha(\nu_{k,1}^{(j)},\nu_{k,2}^{(j)},\dots,\nu_{k,N}^{(j)})$ for all $k=1,\dots,K$ and $j=1,\dots,J$.

An instance of the above procedure is illustrated in the following example.

\begin{example}
\label{ex:motivating}
Suppose that our task consists of $J=4$ jobs. For the $j$-th job, denoted $\mathcal{J}_j$, we need to count $Q=6$ words given by the set $\mathcal{A}^{(j)}=\{\chi_1^{(j)},\dots,\chi_6^{(j)}\}$ in a book consisting of $N=6$ chapters using a cluster of $K=6$ servers.
$\mathcal{J}_j$ is associated with the $j$-th book and its files with the chapters $1^{(j)},\dots,6^{(j)}$. Function $\phi_k^{(j)},k=1,\dots,Q$  (assigned to server $U_k$) counts the word $\chi_k^{(j)}$ of $\mathcal{A}^{(j)}$ in the book indexed by $j$.

We subdivide the original data set of each job into $N=6$ files.
The files of the $j$-th job are partitioned into three batches, namely $\{1^{(j)},2^{(j)}\}$, $\{3^{(j)},4^{(j)}\}$ and $\{5^{(j)},6^{(j)}\}$. Exactly four such batches are stored on each server (\emph{cf.} Fig. \ref{fig:placement_ex1}). The owners of the jobs are specified as follows. 
\begin{equation}
\begin{gathered}
X^{(1)}=\{U_1, U_3, U_5\}, X^{(2)}=\{U_1, U_4, U_6\},\\
X^{(3)}=\{U_2, U_3, U_6\} \text{ and } X^{(4)}=\{U_2, U_4, U_5\}.
\end{gathered}
\label{eq:ex_owners}
\end{equation}

For example, the files of job $\mathcal{J}_1$, $\{1^{(1)},2^{(1)},\dots,6^{(1)}\}$, are stored exclusively on $U_1$, $U_3$ and $U_5$. Specifically, the three batches of the first job are
\begin{equation*}
\mathcal{B}_{[3]}^{(1)}=\{1^{(1)}, 2^{(1)}\},
\mathcal{B}_{[5]}^{(1)}=\{3^{(1)}, 4^{(1)}\},
\mathcal{B}_{[1]}^{(1)}=\{5^{(1)}, 6^{(1)}\}.
\end{equation*}

Then, batch $\mathcal{B}_{[3]}^{(1)}$ is stored on servers $U_1$ and $U_5$, $\mathcal{B}_{[5]}^{(1)}$ on $U_1$ and $U_3$ and, finally, $\mathcal{B}_{[1]}^{(1)}$ on $U_3$ and $U_5$.
Each server locally stores $\mu=\frac{1}{3}$ of all the data sets.

We will clarify the three stages of our proposed Shuffle phase by means of the following example.

\begin{figure}[t]
\centering
\includegraphics[scale=0.64]{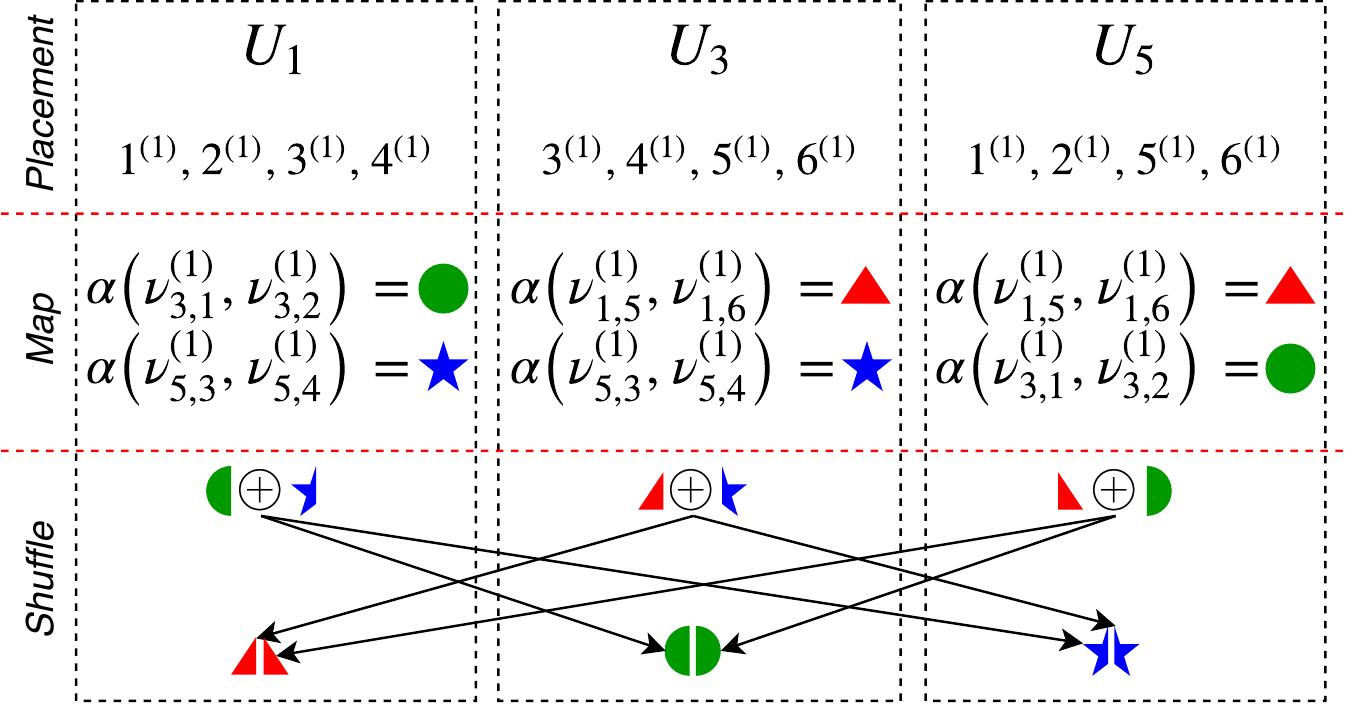}
\caption{Stage 1 coded multicasts among owners of $\mathcal{J}_1$.}
\label{fig:shuffle_ex1}
\vspace{-0.2in}
\end{figure}

\begin{itemize}
\item \textbf{Stage 1}: The owners of each job communicate among themselves during this stage. Let us consider the group of servers $\{U_1,U_3,U_5\}$ which are the owners of $\mathcal{J}_1$, storing $\{1^{(1)},2^{(1)}$, $3^{(1)},4^{(1)}\}$, $\{3^{(1)},4^{(1)},5^{(1)},6^{(1)}\}$ and $\{1^{(1)},2^{(1)},5^{(1)},6^{(1)}\}$, respectively. Based on this allocation policy, server $U_1$ needs $\phi_1^{(1)}$ evaluations of the batch $\{5^{(1)}, 6^{(1)}\}$, i.e.,
$\nu_{1,5}^{(1)} \text{ and } \nu_{1,6}^{(1)}$ for $\mathcal{J}_1$ or simply the aggregate $\alpha(\nu_{1,5}^{(1)}, \nu_{1,6}^{(1)})\triangleq\nu_{1,5}^{(1)}+\nu_{1,6}^{(1)}$ which is the sum of the counts of word $\chi_1^{(1)}$ in files $5^{(1)}$ and $6^{(1)}$. Similarly, $U_3$ needs $\alpha(\nu_{3,1}^{(1)}, \nu_{3,2}^{(1)})$ and $U_5$ needs $\alpha(\nu_{5,3}^{(1)}, \nu_{5,4}^{(1)})$. Next, we refer to Fig. \ref{fig:shuffle_ex1}. The compressed intermediate values are represented by circle/green, star/blue and triangle/red. We further suppose that each value can be split into two packets (represented by the left and right parts of each shape). If $U_1$ transmits left circle XOR left star, then $U_3$ is able to cancel out the star part (since $U_3$ also ``maps" $\{3^{(1)}, 4^{(1)}\}$) and recover the circle part. Similarly, $U_5$ can recover the star part from the same transmission. Each of these transmissions is useful to two servers. We can repeat this process for the remaining jobs. The total number of bits transmitted in this case is therefore $6B$. The incurred communication load is
$L_{\text{stage 1}}=\frac{6B}{JQB}=\frac{1}{4}$. 

\item \textbf{Stage 2}: The groups communicating in this stage consist of both owners and non-owners. The servers recover values of jobs for which they haven't stored any file. Let $G=\{U_1,U_3,U_6\}$. Observe from eq. \eqref{eq:ex_owners} that there is no job common to all three but each subset of two of them shares a batch of a job they commonly own. The remaining server needs an aggregate value of those files. The values that each of $U_1,U_3$ and $U_6$ needs as well as the corresponding transmissions are illustrated in Table \ref{table:ex_stage2}. We denote the $i$-th packet of an aggregate value by $\alpha(\cdot)[i]$. It turns out that there are $4$ possible such groups we can pick. The total load is $L_{\text{stage 2}}=\frac{4\times3\times B/2}{JQB}=\frac{6B}{JQB}=\frac{1}{4}$.

\item \textbf{Stage 3}: Servers recover the remaining intermediate values by receiving unicast transmissions during this last stage. If we consider the same group as in stage 2, i.e., $G=\{U_1,U_3,U_6\}$ then we can see that $U_1$ still misses values $\nu_{1,1}^{(3)}, \nu_{1,2}^{(3)}, \nu_{1,3}^{(3)}$ and $\nu_{1,4}^{(3)}$ of $\mathcal{J}_3$ or simply their aggregate $\alpha(\nu_{1,1}^{(3)}, \nu_{1,2}^{(3)}, \nu_{1,3}^{(3)}, \nu_{1,4}^{(3)})$. Observe that all required files locally reside in the cache of $U_2$ which can transmit the value to $U_1$. For the complete set of unicast transmissions see Table \ref{table:ex_stage2_missing}.
The load turns out to be $L_{\text{stage 3}}=\frac{6\times2\times B}{JQB}=\frac{1}{2}$.
\end{itemize}

The communication load of all stages is then $L_{\text{CAMR}}=1$.
Similarly, the load achieved by the CCDC scheme of \cite{compressed_CDC} for the same storage fraction $\mu=1/3$ is $L_{\text{CCDC}}=1$. Nonetheless, their approach would require a minimum of $J={{6}\choose{3}}=20$ distributed jobs to be executed, i.e., we can achieve the same efficiency on a smaller scale.
\end{example}

\begin{table}[t]
\centering
\caption{Stage 2 transmissions within group $\{U_1,U_3,U_6\}$.}
\label{table:ex_stage2}
\begin{tabular}{|P{0.7cm}||P{4.1cm}|P{1.6cm}|}
\hline
Server&Transmits&Recovers\\
\hline
$U_1$&$\alpha(\nu_{6,3}^{(1)}, \nu_{6,4}^{(1)})[1]\oplus\alpha(\nu_{3,1}^{(2)}, \nu_{3,2}^{(2)})[1]$&$\alpha(\nu_{1,5}^{(3)}, \nu_{1,6}^{(3)})$\\
\hline
$U_3$&$\alpha(\nu_{6,3}^{(1)}, \nu_{6,4}^{(1)})[2]\oplus\alpha(\nu_{1,5}^{(3)}, \nu_{1,6}^{(3)})[1]$&$\alpha(\nu_{3,1}^{(2)}, \nu_{3,2}^{(2)})$\\
\hline
$U_6$&$\alpha(\nu_{3,1}^{(2)}, \nu_{3,2}^{(2)})[2]\oplus\alpha(\nu_{1,5}^{(3)}, \nu_{1,6}^{(3)})[2]$&$\alpha(\nu_{6,3}^{(1)}, \nu_{6,4}^{(1)})$\\
\hline
\end{tabular}
\vspace{-0.2in}
\end{table}

\subsection{Aggregated Multi-Job Communication Load Analysis}
\label{sec:load_camr}
In the first stage, for each of the $J$ jobs, each of the $k$ owners computes one aggregate and is associated with a unique corresponding packet of it, of size $\frac{B}{k-1}$. The communication load is
$$L_{\text{stage 1}}=\frac{Jk\frac{B}{k-1}}{JQB}=\frac{k}{K(k-1)}.$$

The second stage involves the communication within all possible $q^{k-1}(q-1)$ groups that satisfy the desired property. In each case, $k$ servers transmit one value each, of length $\frac{B}{k-1}$ and
$$L_{\text{stage 2}}=\frac{q^{k-1}(q-1)k\frac{B}{k-1}}{JQB}=\frac{(q-1)k}{K(k-1)}.$$

Each server does not own $J-q^{k-2}$ jobs. For each of them, during stage 3, one transmission (of length $B$) from a server in the same parallel class is sufficient. Thus,
$$L_{\text{stage 3}}=\frac{K\left(J-q^{k-2}\right)B}{JQB}=\frac{q-1}{q}.$$

The total load is
\begin{equation}
\label{eq:CAMR_total_load}
L_{\text{CAMR}}=\sum\limits_{i=1}^3L_{\text{stage i}}=\frac{k(q-1)+1}{q(k-1)}.
\end{equation}

\subsection{Comparison With Other Schemes}
\label{sec:load_comparison}
The technique proposed in \cite{compressed_CDC} demonstrates a load of
\begin{equation}
\label{eq:load_ccdc}
L_{\text{CCDC}}=\frac{(1-\mu)(\mu K+1)}{\mu K}.
\end{equation}
for a suitable storage fraction such that $\mu K \in\{1,\dots,K-1\}$. Our storage requirement is equal to $\mu=\frac{k-1}{K}$. For the same storage requirement, eq. \eqref{eq:load_ccdc} yields
\begin{eqnarray*}
L_{\text{CCDC}}&=&\frac{(1-\frac{k-1}{K})(\frac{k-1}{K} K+1)}{\frac{k-1}{K} K}=\frac{k(q-1)+1}{q(k-1)}.
\end{eqnarray*}
We conclude that the loads induced by the two schemes are identical.
However, their approach fundamentally relies on the requirement that the minimum number of jobs to be executed is $J_{\text{CCDC,\ min}}={{K}\choose{\mu K+1}}$. Comparing this value with our requirement for $J_{\text{CAMR}}=q^{k-1}$ and using a known bound for the binomial coefficients, we deduce that \cite{CLRS_book}
\begin{eqnarray*}
J_{\text{CCDC,\ min}}={{K}\choose{\mu K+1}}={{kq}\choose{k}}\labelrel\geq{eq:J_bound}\left(\frac{kq}{k}\right)^k
\labelrel>{eq:CAMR_bound}J_{\text{CAMR,\ min}},
\end{eqnarray*}
where the bound of \eqref{eq:J_bound} is maximum when $q=2$ and becomes stricter for $q>2$; however, for a fixed value of $k$, as $q$ increases the bound of \eqref{eq:CAMR_bound} loosens and it turns out that our requirement for the number of jobs becomes exponentially smaller than that of CCDC (recall that $J_{\text{CAMR,\ min}}=q^{k-1}$).

\begin{table}[t]
\centering
\caption{Needed aggregate values at the end of stage 2.}
\label{table:ex_stage2_missing}
\begin{tabular}{|P{0.7cm}||P{7cm}|}
\hline
Server&Needs\\
\hline
$U_1$&$\alpha(\nu_{1,1}^{(3)}, \nu_{1,2}^{(3)}, \nu_{1,3}^{(3)}, \nu_{1,4}^{(3)})$ and $\alpha(\nu_{1,1}^{(4)}, \nu_{1,2}^{(4)}, \nu_{1,3}^{(4)}, \nu_{1,4}^{(4)})$\\
\hline
$U_2$&$\alpha(\nu_{2,1}^{(1)}, \nu_{2,2}^{(1)}, \nu_{2,3}^{(1)}, \nu_{2,4}^{(1)})$ and $\alpha(\nu_{2,1}^{(2)}, \nu_{2,2}^{(2)}, \nu_{2,3}^{(2)}, \nu_{2,4}^{(2)})$\\
\hline
$U_3$&$\alpha(\nu_{3,3}^{(2)}, \nu_{3,4}^{(2)}, \nu_{3,5}^{(2)}, \nu_{3,6}^{(2)})$ and $\alpha(\nu_{3,3}^{(4)}, \nu_{3,4}^{(4)}, \nu_{3,5}^{(4)}, \nu_{3,6}^{(4)})$\\
\hline
$U_4$&$\alpha(\nu_{4,3}^{(1)}, \nu_{4,4}^{(1)}, \nu_{4,5}^{(1)}, \nu_{4,6}^{(1)})$ and $\alpha(\nu_{4,3}^{(3)}, \nu_{4,4}^{(3)}, \nu_{4,5}^{(3)}, \nu_{4,6}^{(3)})$\\
\hline
$U_5$&$\alpha(\nu_{5,1}^{(2)}, \nu_{5,2}^{(2)}, \nu_{5,5}^{(2)}, \nu_{5,6}^{(2)})$ and $\alpha(\nu_{5,1}^{(3)}, \nu_{5,2}^{(3)}, \nu_{5,5}^{(3)}, \nu_{5,6}^{(3)})$\\
\hline
$U_6$&$\alpha(\nu_{6,1}^{(1)}, \nu_{6,2}^{(1)}, \nu_{6,5}^{(1)}, \nu_{6,6}^{(1)})$ and $\alpha(\nu_{6,1}^{(4)}, \nu_{6,2}^{(4)}, \nu_{6,5}^{(4)}, \nu_{6,6}^{(4)})$\\
\hline
\end{tabular}
\vspace{-0.2in}
\end{table}

\begin{table*}[t]
\centering
\caption{Time for computing $512$ products $\mathbf{Ab}$, $m=234000$, $n=100$ on $K=20$ servers.}
\resizebox{2\columnwidth}{!}{
\begin{tabular}{|P{2.3cm}||P{1.2cm}|P{0.7cm}|P{1cm}|P{1cm}|P{1.1cm}|P{1.1cm}|P{1.6cm}|P{1.2cm}|P{1cm}|}
\hline
&CodeGen&Map&Encode&Shuffle&Decode&Reduce&Total Time&Speedup&Rate\\
&(sec.)&(sec.)&(sec.)&(sec.)&(sec.)&(sec.)&(sec.)&&(Mbps.)\\
\hline
Uncoded: &-&1.01&-&408.11&-&0.17&409.29&&169.10\\
Proposed: $k=10$&29.63&8.77&0.41&52.74&3.43&0.04&95.02&$4.31\times$&90.85\\
\hline
\end{tabular}
}\\
\label{table:camr_k10_tests}
\vspace{-0.2in}
\end{table*}

\subsection{Distributed Matrix-Vector Multiplication}
\label{sec:mat_vec_multiplication}
Performing large matrix-vector multiplications is a key building block of several machine learning algorithms. For instance, during the forward propagation in deep neural networks \cite{goodfellow2016deep} the output of the layer is the result of multiplying the matrix of the input data set with the weight vector. In what follows, we formulate the matrix-vector product as a MapReduce operation and compare our algorithm against the baseline method for the case when we have to simultaneously execute multiple such multiplications. Existing work on the multi-job case does not include practical experiments. Thus, we cannot compare with other schemes that examine the computation-communication trade-off on multiple jobs. Nevertheless, we believe that our experiments provide a good demonstration of potential benefits of these operations on a large scale.

Suppose that we want to compute $\mathbf{Ab}$ for a matrix $\mathbf{A}$ (size $m\times Jn$) with a vector $\mathbf{b}$ (size $Jn\times 1$) in a distributed manner on $K$ servers. We assume $K|m$. We initially split it column-wise into $J$ blocks $\mathbf{A}^{(1)}, \dots, \mathbf{A}^{(J)}$. Each block is associated with one job. Specifically, the job indexed with $j$ involves multiplying $\mathbf{A}^{(j)}$ (size $m\times n$) with a vector $\mathbf{b}^{(j)}$ (size $n\times 1$).

We will begin by explaining our model for the baseline approach. First, we further partition each $\mathbf{A}^{(j)}$ into $qk$ block matrices as follows
\begin{equation}
\label{eq:A_split}
\mathbf{A}^{(j)}=
\left[
\begin{array}{c|c|c}
\mathbf{A}_{11}^{(j)} & \dots & \mathbf{A}_{1k}^{(j)} \\
\hline
\vdots & \ddots & \vdots \\
\hline
\mathbf{A}_{q1}^{(j)} & \dots & \mathbf{A}_{qk}^{(j)} \\
\end{array}
\right].
\end{equation}

The corresponding decomposition of $\mathbf{b}^{(j)}$ into blocks is as follows
\begin{equation}
\label{eq:b_split}
\mathbf{b}^{(j)}=
\begin{bmatrix}
b_{1}^{(j)}&\cdots&b_{n/k}^{(j)}|\cdots|b_{n-n/k+1}^{(j)}\cdots b_{n}^{(j)}\\
\end{bmatrix}^T.
\end{equation}
Each server stores and computes the product of exactly one block of $\mathbf{A}^{(j)}, \forall j\in[J]$ (there are $K=kq$ of them) with the appropriate subvector of $\mathbf{b}^{(j)}$, during the Map phase. Our Reduce policy is that each server will compute a subset of the rows of $\mathbf{c}^{(j)}=\mathbf{A}^{(j)}\mathbf{b}^{(j)}$ after processing at its end. Specifically, server $U_i$ is assigned to compute the rows $\{(i-1)m/K, \dots, im/K\}$ of $\mathbf{c}^{(j)}$ (note that we assume that $K|m$). All $K$ reducers receive $k-1$ products (size $m/K\times 1$) for each job by servers mapping the same block-row of $\mathbf{A}^{(j)}$ and sum these results row-wise before transmitting them to the master. The master machine concatenates them and constructs the final result.


Let $c(m,n,k)$ be the cost of multiplying two matrices of dimension $m\times n$ and $n\times k$. Then, the computation cost for each server is $M_{\text{uncoded}}=J\cdot c(m/q,n/k,1)$.

The communication load is $L_{\text{uncoded}}^{\text{mult}}=\frac{JK(k-1)B}{JQB}=k-1$ where, based on our prior notation, $B=\frac{m}{K}T$ and $T$ is the number of bits used to represent a single entry of a matrix, i.e., each transmission is the equivalent of a ``compressed" intermediate value (a column in this case).

We now formulate our CAMR scheme for this problem. In this case, we split $\mathbf{A}^{(j)}$ into $k$ block-columns as $$\mathbf{A}^{(j)}=\begin{bmatrix}\mathbf{A}_{1}^{(j)}, & \dots, & \mathbf{A}_{k}^{(j)}\end{bmatrix}.$$ For each job (point), we pick $k$ owners (blocks) based on a SPC-$(k,k-1)$ code that store a part of $\mathbf{A}^{(j)}$ and $\mathbf{b}^{(j)}$ (the splitting of $\mathbf{b}^{(j)}$ is the same as in eq. \ref{eq:b_split}). Specifically, each owner stores a different set of $k-1$ block-columns (batches) of $\mathbf{A}^{(j)}$ and the corresponding parts of $\mathbf{b}^{(j)}$. It computes all these products during the Map phase. The non-owners do not store any part of these matrices. The Reduce policy also remains the same as in the baseline method.

The computation cost per server is $M_{\text{CAMR}}=q^{k-2}(k-1)\cdot c(m,n/k,1)$. The communication load has been computed in Section \ref{sec:load_camr}, eq. \eqref{eq:CAMR_total_load}.

In theory, CAMR requires a computation overhead of
\begin{equation*}
\frac{M_{\text{CAMR}}}{M_{\text{uncoded}}}=\frac{q^{k-2}(k-1)\cdot c(m,n/k,1)}{J\cdot c(m/q,n/k,1)}=k-1.
\end{equation*}
The theoretical gain we would expect in the Shuffle phase is
\begin{equation*}
\frac{L_{\text{uncoded}}^{\text{mult}}}{L_{\text{CAMR}}}=\frac{(k-1)^2q}{k(q-1)+1}.
\end{equation*}

\subsection{Matrix-vector Multiplication Experimental Results and Discussion}
We serially ran multiple matrix-vector products on Amazon EC2 clusters. The instance type used is x1e.2xlarge for the master machine and r4.2xlarge for the workers. Our code is available online \cite{compressed_repo}. The master machine decomposes each input matrix and the corresponding vector and sends them to the appropriate worker nodes.

Table \ref{table:camr_k10_tests} summarizes the results for our use case. The impact of the Shuffle phase on the total execution time seems to be greater than in the case of TeraSort and our scheme reduces the overall time by up to $4.31\times$. In theory, our scheme requires a computation overhead of $9$. Indeed, based on Table \ref{table:camr_k10_tests} the Map phase for our scheme is $\frac{8.77}{1.01}\approx8.68$ times more expensive than that of the uncoded. The gain we would expect in the Shuffle phase for the values of the parameters is approximately $14.73$. In practice, we have achieved a gain of $\frac{408.11}{52.74}\approx7.74$. Nevertheless, if we consider equal transmission rates in both cases that speedup would be $\frac{169.1}{90.85}7.74=14.41$ which is very close to the prediction.

\section{Conclusions And Future Work}
In this work we presented a distributed computing protocol by leveraging the properties of resolvable designs. These designs can be generated from single parity-check codes. Our techniques apply for the execution of a single job with arbitrary functions and a multi-job scenario where the functions can be aggregated. Prior work has identified and proposed techniques for exploring these tradeoffs for both problems. However, in both cases those techniques require certain problem dimensions to be very large in the problem parameters. Specifically, in the single-job case, they require a large number of files, whereas in the multi-job case they require a large number of jobs. In practical scenarios, this is a serious issue and adversely affects the job execution times. Our proposed approaches work with significantly smaller number of subfiles (single-job) and jobs (multi-job), respectively. We theoretically analyze the performance of our schemes and also present exhaustive experiments on Amazon EC2 platforms that confirm the performance advantages of our methods.

We point out that our number of subfiles is still exponential in the problem parameters but with a much smaller exponent. We emphasize that it remains well within the limits of popular message-passing protocols such as Open MPI for many practical scenarios. Reducing this number further while continuing to have a low communication load is an interesting direction for future work. Our multi-job scheme (and prior work \cite{compressed_CDC}) does not handle precedence constraints or a redundant Reduce function assignment to the workers that naturally arise in some MapReduce problems. Adapting our work to take these into account would be another avenue for future work.

%
%
%
%
%

\appendix

\subsection*{Proof of Lemma \ref{lemma:shufffling_lemma}}
We shall refer to Algorithm \ref{alg:shuffling_lemma} in order to show that each server in $G$ can recover its missing data chunk. For a group of servers $G=\{U_1,\dots,U_k\}$ and the packets of the chunk $\mathcal{D}_{[j]}$, i.e., $\mathcal{C}=\{\mathcal{D}_{[j]}[1],\dots,\mathcal{D}_{[j]}[k-1]\}$ consider a complete graph with the following set of vertices $\{G\setminus\{U_j\}, \mathcal{C}\}$; the matching $H^{[j]}$ described in Algorithm \ref{alg:shuffling_lemma} will be defined based upon this graph, i.e., each vertex in $\{G\setminus\{U_j\}\}$ will be matched to a distinct vertex in $\mathcal{C}$. Fix a pair of servers $\{U_m,U_k\}\subset G$ and the packet $\Delta_m$ transmitted from $U_m$ to $U_k$. By canceling out all terms of $\Delta_m$ 
that $U_k$ locally stores, it can recover the remaining term, i.e., $H^{[k]}(U_m)$ (note that $U_k$ participates in exactly $k-1$ such matchings, i.e., in $H^{[j]}, \forall j\neq k$). Keeping $U_{k}$ fixed, we repeat this process for every possible server $U_m\in G\setminus\{U_{k}\}$. Since each of them is associated with a distinct packet of $\mathcal{D}_{[k]}$ it follows that by receiving the $k-1$ packets
$$\{\Delta_m: U_m\in G\setminus\{U_{k}\}\},$$
$U_k$ can recover the following packets
$$\{\mathcal{D}_{[k]}[i]:U_i\in G\setminus\{U_{k}\}\}.$$
Subsequently, $U_{k}$ concatenates them in order to recover $\mathcal{D}_{[k]}$. Since this proof holds independent of the choice of $U_m$, we have shown that all servers can recover their missing chunks. To conclude the argument, we note that since each chunk is assumed to be of size $B$ bits and it was split into $k-1$ packets of size $B/(k-1)$, the total amount of transmitted data is $Bk/(k-1)$.


\section*{Details of TeraSort Implementation}
We have implemented TeraSort on Amazon EC2 clusters  using our proposed approach. The implementation was performed in C++ using the Open MPI library for communication among the processes of the master and the servers. Our code builds on \cite{USCTS} and comparisons with the uncoded case and the approach in \cite{songze_terasort} have been made.

TeraSort is a popular benchmark that measures the time to sort a big amount of randomly generated data on a cluster. The data set in TeraSort is such that each line of the file is a key-value (KV) pair typically consisting of an integer key and an arbitrary string value. The sorting is done based on the key. It is not too hard to see that this KV formulation can be put in on-to-one correspondence with the formulation in terms of Map and Reduce functions ({\it cf.} Section \ref{sec:formulation_one_job}).

\subsection{Amazon EC2 cluster configuration}
We used Amazon EC2 instances among which one served as a master and the rest of them as servers (servers). The instance type used is r3.large for the master machine and m3.large for the servers. After placing the files to the carefully chosen servers we also impose a limit of 100Mbps for both incoming and outgoing traffic of all servers\footnote{This serves the purpose of alleviating bursty TCP transmissions.}.

\subsection{Data set description}
For the TeraSort experiments we generated 12GB of total data. Each row of the file holds a 10-byte key and its corresponding 90-byte value. The TeraGen utility of Hadoop distribution was used to randomly generate this data. The KV pairs are lexicographically sorted with respect to the ASCII code of their keys where the leftmost and the rightmost byte are the most and the least significant byte, respectively.

\subsection{Platform and code implementation description}
Our source code is available at \cite{KKCT}. The master machine is responsible for placing the files in the local drives of the servers and deciding the reducer responsibilities for each server. It also initiates the MPI program to all servers. From this point onwards, the master will only take time measurements from the servers.


The overall sequence of steps in processing a given job are: CodeGen $\rightarrow$ Map $\rightarrow$ Pack/Encode $\rightarrow$ Shuffle $\rightarrow$ Unpack/Decode $\rightarrow$ Reduce. We explain these steps below.

\begin{itemize}
\item \emph{Code generation}: All nodes (including the master) start by generating the resolvable design based on our choice of the parameters $q$ and $k$. Next, the data set is split into $N$ files by the master and the appropriate files are transmitted to each server based on Step \ref{alg:file_match} of Algorithm \ref{Alg:protocol}.
The master also broadcasts the keys that describe the Reduce assignment.
\item \emph{Map.} For each file $w_a$ that server $B_{i,l}$ has in its block, it will compute $\{\nu_{1,a},\dots,\nu_{Q,a}\}$ during the Map phase.
\item \emph{Pack/Encode.} For the uncoded implementation, we use the Pack operation. The Pack stage stores all intermediate values that will be sent to the same reducer in a continuous memory array so that a single TCP connection for each sender/receiver pair suffices (which may transmit multiple KV pairs) when $\texttt{MPI\_Send}$ is called\footnote{In the shuffling phase of the uncoded case, each server unicasts data to a single receiver at any particular time, which is exactly the purpose of $\texttt{MPI\_Send}$ call.}. In the coded implementation encoded packets are created from the mapped data as described in Algorithms  \ref{alg:shuffling_lemma}, \ref{Alg:protocol}.
\item \emph{Shuffle.} For each shuffling group $G$ a server belongs to, it will broadcast an appropriate encoded packet to the rest of the group.
\item \emph{Unpack/Decode.} In the uncoded implementation we use the Unpack operation which simply deserializes the received data to a list of KV pairs.
In the coded implementation the intermediate values are decoded locally on each server from the received data.
\item \emph{Reduce.} The Reduce function is applied on the unpacked/decoded data.
\end{itemize}

\bibliographystyle{IEEEtran}
\bibliography{./citations}

\begin{thebibliography}{10}
\providecommand{\url}[1]{#1}
\csname url@samestyle\endcsname
\providecommand{\newblock}{\relax}
\providecommand{\bibinfo}[2]{#2}
\providecommand{\BIBentrySTDinterwordspacing}{\spaceskip=0pt\relax}
\providecommand{\BIBentryALTinterwordstretchfactor}{4}
\providecommand{\BIBentryALTinterwordspacing}{\spaceskip=\fontdimen2\font plus
\BIBentryALTinterwordstretchfactor\fontdimen3\font minus
  \fontdimen4\font\relax}
\providecommand{\BIBforeignlanguage}[2]{{%
\expandafter\ifx\csname l@#1\endcsname\relax
\typeout{** WARNING: IEEEtran.bst: No hyphenation pattern has been}%
\typeout{** loaded for the language `#1'. Using the pattern for}%
\typeout{** the default language instead.}%
\else
\language=\csname l@#1\endcsname
\fi
#2}}
\providecommand{\BIBdecl}{\relax}
\BIBdecl

\bibitem{compressed_CDC}
S.~Li, M.~A. Maddah{-}Ali, and A.~S. Avestimehr, ``Compressed coded distributed
  computing,'' in \emph{{IEEE} International Symposium on Information Theory
  ({ISIT})}, June 2018, pp. 2032--2036.

\bibitem{he_resnet}
K.~He, X.~Zhang, S.~Ren, and J.~Sun, ``Deep residual learning for image
  recognition,'' in \emph{{IEEE} Conference on Computer Vision and Pattern
  Recognition ({CVPR})}, June 2016, pp. 770--778.

\bibitem{saurav_g_analytics}
S.~Prakash, A.~Reisizadeh, R.~Pedarsani, and A.~S. Avestimehr, ``Coded
  computing for distributed graph analytics,'' in \emph{{IEEE} International
  Symposium on Information Theory ({ISIT})}, June 2018, pp. 1221--1225.

\bibitem{Chowdhury_etal11}
M.~Chowdhury, M.~Zaharia, J.~Ma, M.~I. Jordan, and I.~Stoica, ``Managing data
  transfers in computer clusters with orchestra,'' \emph{{ACM SIGCOMM} Computer
  Communication Review}, vol.~41, no.~4, pp. 98--109, August 2011.

\bibitem{GuoRZ13}
Y.~Guo, J.~Rao, and X.~Zhou, ``ishuffle: Improving hadoop performance with
  shuffle-on-write,'' in \emph{10th International Conference on Autonomic
  Computing ({ICAC})}, June 2013, pp. 107--117.

\bibitem{Verma2013}
A.~Verma, B.~Cho, N.~Zea, I.~Gupta, and R.~H. Campbell, ``Breaking the
  mapreduce stage barrier,'' \emph{Cluster Computing}, vol.~16, no.~1, pp.
  191--206, March 2013.

\bibitem{EZZELDIN_KARMOOSE_FRAGOULI_2017}
Y.~H. Ezzeldin, M.~Karmoose, and C.~Fragouli, ``Communication vs distributed
  computation: An alternative trade-off curve,'' in \emph{2017 IEEE Information
  Theory Workshop (ITW)}, November 2017, pp. 279--283.

\bibitem{Chen_graph14}
R.~Chen, X.~Ding, P.~Wang, H.~Chen, B.~Zang, and H.~Guan, ``Computation and
  communication efficient graph processing with distributed immutable view,''
  in \emph{23rd International Symposium on High-performance Parallel and
  Distributed Computing ({HPDC})}, June 2014, pp. 215--226.

\bibitem{LiMA16}
S.~Li, M.~A. Maddah-Ali, Q.~Yu, and A.~S. Avestimehr, ``A fundamental tradeoff
  between computation and communication in distributed computing,'' \emph{IEEE
  Transactions on Information Theory}, vol.~64, no.~1, pp. 109--128, January
  2018.

\bibitem{Ahmad_2014}
F.~Ahmad, S.~T. Chakradhar, A.~Raghunathan, and T.~N. Vijaykumar,
  ``Shufflewatcher: Shuffle-aware scheduling in multi-tenant mapreduce
  clusters,'' in \emph{USENIX Annual Technical Conference (ATC)}, June 2014,
  pp. 1--13.

\bibitem{Cao_2016}
X.~Cao, K.~K. Panchputre, and D.~H.-C. Du, ``Accelerating data shuffling in
  mapreduce framework with a scale-up numa computing architecture,'' in
  \emph{24th High Performance Computing Symposium}, April 2016, pp. 17:1--17:8.

\bibitem{Wang_2013}
C.~Wang, Y.~Qin, Z.~Huang, Y.~Peng, D.~Li, and H.~Li, ``Optas: Optimal data
  placement in mapreduce,'' in \emph{International Conference on Parallel and
  Distributed Systems}, December 2013, pp. 315--322.

\bibitem{LiMAAllerton15}
S.~Li, M.~A. Maddah-Ali, and A.~S. Avestimehr, ``Coded mapreduce,'' in
  \emph{53rd Annual Allerton Conference on Communication, Control, and
  Computing (Allerton)}, 2015, pp. 964--971.

\bibitem{songze_terasort}
S.~Li, S.~Supittayapornpong, M.~A. Maddah-Ali, and S.~Avestimehr, ``Coded
  terasort,'' in \emph{{IEEE} International Parallel and Distributed Processing
  Symposium Workshops ({IPDPSW})}, May 2017, pp. 389--398.

\bibitem{tandon_gradient}
R.~Tandon, Q.~Lei, A.~G. Dimakis, and N.~Karampatziakis, ``Gradient coding:
  Avoiding stragglers in distributed learning,'' in \emph{34th International
  Conference on Machine Learning ({ICML})}, August 2017, pp. 3368--3376.

\bibitem{Dally_nips_tutorial}
W.~Dally, ``High-performance hardware for machine learning,'' in \emph{29th
  Conference on Neural Information Processing Systems ({NIPS}) Tutorial},
  December 2015.

\bibitem{DeanG08}
J.~Dean and S.~Ghemawat, ``Mapreduce: Simplified data processing on large
  clusters,'' \emph{Communications of the ACM}, vol.~51, no.~1, pp. 107--113,
  January 2008.

\bibitem{WOOLSEY_CHEN_MINGYUE_2018}
N.~Woolsey, R.~Chen, and M.~Ji, ``A new combinatorial design of coded
  distributed computing,'' in \emph{2018 IEEE International Symposium on
  Information Theory (ISIT)}, June 2018, pp. 726--730.

\bibitem{Song_Srinivasavaradhan_Fragouli_2017}
L.~Song, S.~R. Srinivasavaradhan, and C.~Fragouli, ``The benefit of being
  flexible in distributed computation,'' in \emph{2017 IEEE Information Theory
  Workshop (ITW)}, November 2017, pp. 289--293.

\bibitem{konstantinidis_ramamoorthy_globecom}
K.~Konstantinidis and A.~Ramamoorthy, ``Leveraging coding techniques for
  speeding up distributed computing,'' in \emph{{IEEE} Global Communications
  Conference ({GLOBECOM})}, December 2018, pp. 1--6.

\bibitem{kostasR19}
------, ``{CAMR: Coded Aggregated MapReduce},'' in \emph{{IEEE} International
  Symposium on Information Theory ({ISIT})}, June 2019.

\bibitem{DRSCDCA}
D.~R. Stinson, \emph{Combinatorial Designs: Constructions and Analysis}.\hskip
  1em plus 0.5em minus 0.4em\relax Springer, 2004.

\bibitem{KKCT}
\BIBentryALTinterwordspacing
``{SPC Coded TeraSort repository}.'' [Online]. Available:
  \url{https://bitbucket.org/kkonstantinidis/codedterasort}
\BIBentrySTDinterwordspacing

\bibitem{compressed_repo}
\BIBentryALTinterwordspacing
``{Aggregated MapReduce code repository}.'' [Online]. Available:
  \url{https://bitbucket.org/kkonstantinidis/camrmm}
\BIBentrySTDinterwordspacing

\bibitem{lincostello}
S.~Lin and D.~J. Costello, \emph{Error Control Coding, 2nd Ed.}\hskip 1em plus
  0.5em minus 0.4em\relax Prentice Hall, 2004.

\bibitem{TangR18}
L.~Tang and A.~Ramamoorthy, ``Coded caching schemes with reduced
  subpacketization from linear block codes,'' \emph{{IEEE} Transactions on
  Information Theory}, vol.~64, no.~4, pp. 3099--3120, April 2018.

\bibitem{USCTS}
\BIBentryALTinterwordspacing
``{Repository of TeraSort for prior implementation}.'' [Online]. Available:
  \url{https://github.com/AvestimehrResearchGroup/Coded-TeraSort/tree/IgnoreMemoryTime}
\BIBentrySTDinterwordspacing

\bibitem{Kwon:2012:SMS:2213836.2213840}
Y.~Kwon, M.~Balazinska, B.~Howe, and J.~Rolia, ``Skewtune: Mitigating skew in
  mapreduce applications,'' in \emph{Proceedings of the 2012 ACM SIGMOD
  International Conference on Management of Data}, May 2012, pp. 25--36.

\bibitem{DBLP:journals/debu/KwonRBH13}
Y.~Kwon, K.~Ren, M.~Balazinska, and B.~Howe, ``Managing skew in hadoop,''
  \emph{{IEEE} Data Eng. Bull.}, vol.~36, no.~1, pp. 24--33, March 2013.

\bibitem{Hoefler_MPI_Bcast}
T.~{Hoefler}, C.~{Siebert}, and W.~{Rehm}, ``A practically constant-time mpi
  broadcast algorithm for large-scale infiniband clusters with multicast,'' in
  \emph{2007 IEEE International Parallel and Distributed Processing Symposium},
  March 2007, pp. 1--8.

\bibitem{Sack2010}
P.~Sack and W.~Gropp, ``A scalable mpi{\_}comm{\_}split algorithm for exascale
  computing,'' in \emph{Recent Advances in the Message Passing Interface}, vol.
  6305, September 2010, pp. 1--10.

\bibitem{CLRS_book}
T.~H. Cormen, C.~E. Leiserson, R.~L. Rivest, and C.~Stein, \emph{Introduction
  to Algorithms, Third Edition}, 2009.

\bibitem{goodfellow2016deep}
I.~Goodfellow, Y.~Bengio, A.~Courville, and Y.~Bengio, \emph{Deep
  learning}.\hskip 1em plus 0.5em minus 0.4em\relax MIT Press, 2016.

\end{thebibliography}

\end{document}